\documentclass[useAMS,usenatbib]{mn2e}
\usepackage{epsfig}
\usepackage{amsmath}

\title[Collisionless dynamics in Globular Clusters]
  {Collisionless dynamics in Globular Clusters}
\author[L.L.R. Williams et al.]
  {Liliya L.~R. Williams,$^1$\thanks{email: llrw@astro.umn.edu}
   Eric I. Barnes,$^2$\thanks{email: barnes.eric@uwlax.edu}
   Jens Hjorth,$^3$\thanks{email: jens@dark-cosmology.dk}\\
  $^1$School of Physics and Astronomy, University of Minnesota, 
      116 Church Street SE, Minneapolis, MN 55455, USA\\
  $^2$Department of Physics, University of Wisconsin --- La Crosse,
      La Crosse, WI 54601, USA\\
  $^3$Dark Cosmology Centre, Niels Bohr Institute, 
University of Copenhagen, Juliane Maries Vej 30, DK-2100 Copenhagen \O, Denmark}

\pagerange{\pageref{firstpage}--\pageref{lastpage}} \pubyear{2012}

\begin{document}

\label{firstpage}

\maketitle

\begin{abstract}
Since globular clusters (GCs) are old, low-$N$ systems their dynamics
is widely believed to be fully dominated by collisional two-body
processes, and their surface brightness profiles are fit by King
models. However, for many GCs, especially those with HST-resolved
central regions, and `extra-tidal' features, King models provide poor
fits. We suggest that this is partly because collisionless dynamics is also
important and contribute to shaping the cluster properties. We show
using time-scale and length-scale arguments that except for the very
centers of clusters, collisionless dynamics should be more important
than collisional. We then fit 38 GCs analyzed by \cite{ng06} with
(collisional) King and (collisionless) DARKexp models over the full
available radial range, and find that the latter provide a better fit
to 29 GCs; for six of these the fit is at least $\sim 5\times$ better
in term of $rms$. DARKexp models are theoretically derived maximum
entropy equilibrium states of self-gravitating collisionless systems
and have already been shown to fit the results of dark matter N-body
simulations. (We do not attempt fits with ad hoc fitting functions.)
\end{abstract}

\begin{keywords}
\end{keywords}







\section{Introduction}

Globular Clusters (GCs) are old and contain a relatively small number
($\sim 10^5-10^6$) of stars, as compared to galaxies, and are therefore 
believed to have relaxed through two-body
encounters, both strong and weak. In a Hubble time some GCs would have
formed many compact binaries, reached core collapse, and possibly
undergone several gravothermal oscillations \citep{ehi87,mh97}.

The theoretical treatment of this gravitational evolution problem is
not straightforward.  
While it is possible to describe realistic thermal equilibria for
self-gravitating systems (see \S~\ref{mo}), the negative heat capacity
of such systems implies that these equilibria are not stable.  Any
reduction in total energy leads to an increase in the kinetic
temperature of the system, however, it is reasonable to assume that
if GCs evolve slowly enough they go from one quasi-equilibrium state
to another.  These quasi-equilibrium states are often described by King
equilibrium models, which attempt to take into account two key
characteristics of GCs: relaxation through two-body encounters, and
Galactic tidal field. To incorporate these characteristics, \cite{k66}
starts with the Maxwell-Boltzmann distribution function, $f(E)\propto
\exp([\Phi_0-E]/\sigma^2)$, which corresponds to the isothermal sphere
of central potential $\Phi_0$ and constant velocity dispersion
$\sigma$, and truncates it to imitate the result of spatial tidal
stripping experienced by systems in the Galactic tidal field,
$f(E)\propto \exp([\Phi_0-E]/\sigma^2)-1$;  see also \cite{m63}.
King-Michie models extend isotropic King models to include velocity
dispersion anisotropy.

It is generally believed that GCs are well-fit by King, or King-Michie
models \citep{ehi87,mh97}. However, deviations from King models have
always been known to exist, both at small and large radii. At small
radii, the central density profiles, especially in HST data, sometimes
lack the flat density cores required by King models but instead
exhibit a range of power-law cusps \citep{ng06}. At large radii, many
clusters are better fit with power-law density profiles that remain
shallower than King to the last observable point \citep{mm05,cb11}.
Other GCs appear to contain `extra-tidal' stars, which make the surface
brightness profiles more extended than King models. 

Because some GCs are known to show mass segregation, many authors argue
that GCs should be fit with multiple King models, each representing
some narrow stellar mass range.  Profiles consisting of several (and
as many as 10) mass components have been presented \citep{df76}.
Several authors have also pointed out that models other than King can
provide more conforming fits, like Wilson models \citep{w75,mm05}, or
the power law models \citep{cb11}.  The main drawback of using the
latter two models is that they are ad hoc fitting functions.

We are motivated by the widespread deviations between observations
and King models and the wide variety of proposed fixes to ask whether
King is really the best basic model for GCs.  More specifically, we ask 
if the two-body encounters that involve one or few stars are as dominant 
a feature as it is generally believed, or if collective collisionless 
effects on scales of hundreds or thousands of stars are also
important.

There is a possible indication of collisionless dynamics in at least 
one cluster. WLM-1, studied in detail by \cite{s06}, has an ellipticity
of $e=0.17$ which they conclude is not the result of rotation or
Galactic tides.  Instead, the shape is most likely due to anisotropic
velocity dispersions, as occurs in many ellipticals. Two-body effects
would have wiped out differences in velocity dispersion in different
directions, which implies that two-body relaxation has not been very
efficient in this GC.

Violent relaxation, a collisionless process, is brought about by fluctuations 
in the large-scale, or global potential of the system.  The difference from 
collisional relaxation is that particles---stars---exchange energy with the
large-scale potential, and not directly among themselves. In the case
of galaxies, the potential fluctuations are usually the consequence of
collapse and subsequent oscillations that occur during formation. In
the case of GCs, the time since initial collapse is generally longer
than the two-body relaxation time-scale, and so additional causes for
global potential fluctuations might be necessary if collisionless
relaxation is to be relevant for present-day GCs, such as GC's elliptical 
orbits in the Galaxy, disk passage, Galactic tides, and core 
collapse and bounce. For example, from the proper motion measurements 
of NCG 6397 \cite{kal07} conclude that the cluster has made frequent passages 
through the Galactic disk.

Mixing is another aspect of collisionless relaxation; it does not result in
energy changes of particles, but is also thought to be important in the formation
of galaxies \citep{m05}. It is difficult to assign a specific mechanism responsiblity,
because even though the importance of potential fluctuations and mixing
as drivers of collisionless relaxation is well established, the full
understanding of relaxation and its detailed mechanics is still
lacking. So instead of investigating the relaxation mechanisms, we take a
different approach. As we describe in Section~\ref{De}, it is possible
to derive the final state attained by the system after collisionless 
relaxation is complete, with no regard for the dynamics that lead to it. 
The final relaxed state can be described by the mass density distribution,
which can then be compared to that of GCs.

Therefore, in the present paper we do not undertake any dynamical
experiments.  We fit GC surface brightness (SB) profiles from the
literature, with two functions: King and DARKexp, which represent
fully collisional and fully collisionless systems, respectively.
Section~\ref{mo} shows that each model has been derived from maximum
entropy arguments implying that these are effective quasi-equilibrium
models for the final state of the two types of systems. We do not fit
ad hoc or loosely motivated functions.

\section{Assessing collisionality of Globular Clusters}\label{coll}

It is well established that two-body, or collisional processes are
responsible for many properties of GCs, like the presence and hardening
of binaries, mass segregation, and possibly energy transport and 
core collapse \citep{mh97}.

Here we ask, how important are collisionless effects in Globular Clusters?  
One way to address the question is to estimate relevant time-scales. 
The two-body relaxation time-scale for GCs is shorter than the Hubble time.
However, one must also compare it to the collisionless relaxation
time-scale which proceeds on dynamical, or approximately 
crossing time-scales \citep{bt87,lb67}.

The two-body relaxation time-scale is of the order,
$t_{2bd}=(r/v)N/(8\ln N)$, where $r$ is the size of the system, $v$ is
the characteristic velocity of the particles, and $N$ is the number of
particles. The dynamical time-scale is, $t_{dyn}=2\pi (r/v_c)$, where
$v_c$ is the circular velocity and is of the same order as $v$. The
ratio of the two, $t_{dyn}/t_{2bd}=16\pi (\ln N)/N$ is $2.3$ for
$N=100$ and $6\times10^{-3}$ for $N=10^5$.  This difference suggests that
collisionless relaxation will drive the overall dynamical evolution of
a GC, but interior to radii where $N\sim 100$, the two-body processes
become more important for the evolution.

Another way to access the relevance of collisionless effects as a
function of distance $r$ from cluster center is to ask 
whether the cluster's acceleration field is grainy or smooth on the scale 
of individual stars. For this, one needs to estimate
two length-scales. The first, $r_{eq}$ is the distance from the star
where the typical acceleration induced by that star equals the
acceleration due to the global potential; 
\begin{equation}
\frac{GM(<r)}{r^2}=\frac{Gm}{r_{eq}^2}.
\end{equation}
Here, the total mass of the cluster is related to the mass of a single
star, and the total number of stars, $M_{tot}=mN_{tot}$. The other
length-scale, $r_{sep}$, is the typical separation between stars at
radius $r$, $r_{sep}=[(4\pi/3)\rho(r)/m]^{-1/3}$.  The ratio of the
two is
\begin{equation}
\frac{r_{sep}}{r_{eq}}=\Bigl(\frac{3 M_{tot}}{4\pi \rho(r)r^3}\Bigr)^{1/3}
                       \Bigl(\frac{M(<r)}{M_{tot}}\Bigr)^{1/2}
		       N_{tot}^{1/6}
\end{equation}

If $r_{sep}\ll r_{eq}$, as one would find near the center of a GC, then the 
acceleration field and the potential are grainy, so the situation is collisional. 

Note that because this calculation assesses the
graininess of the acceleration field (at distance $r$), it includes all
types of collisional encounters---the strong ones with nearby stars
and the weak ones with more distant stars.  

If $r_{sep}\gg r_{eq}$, for example, in the outer regions of a GC, then the 
acceleration field and the potential are smooth, and the situation is collisionless. 

\begin{figure*}
  \begin{center}
    \leavevmode
      \epsfxsize15cm\epsfbox{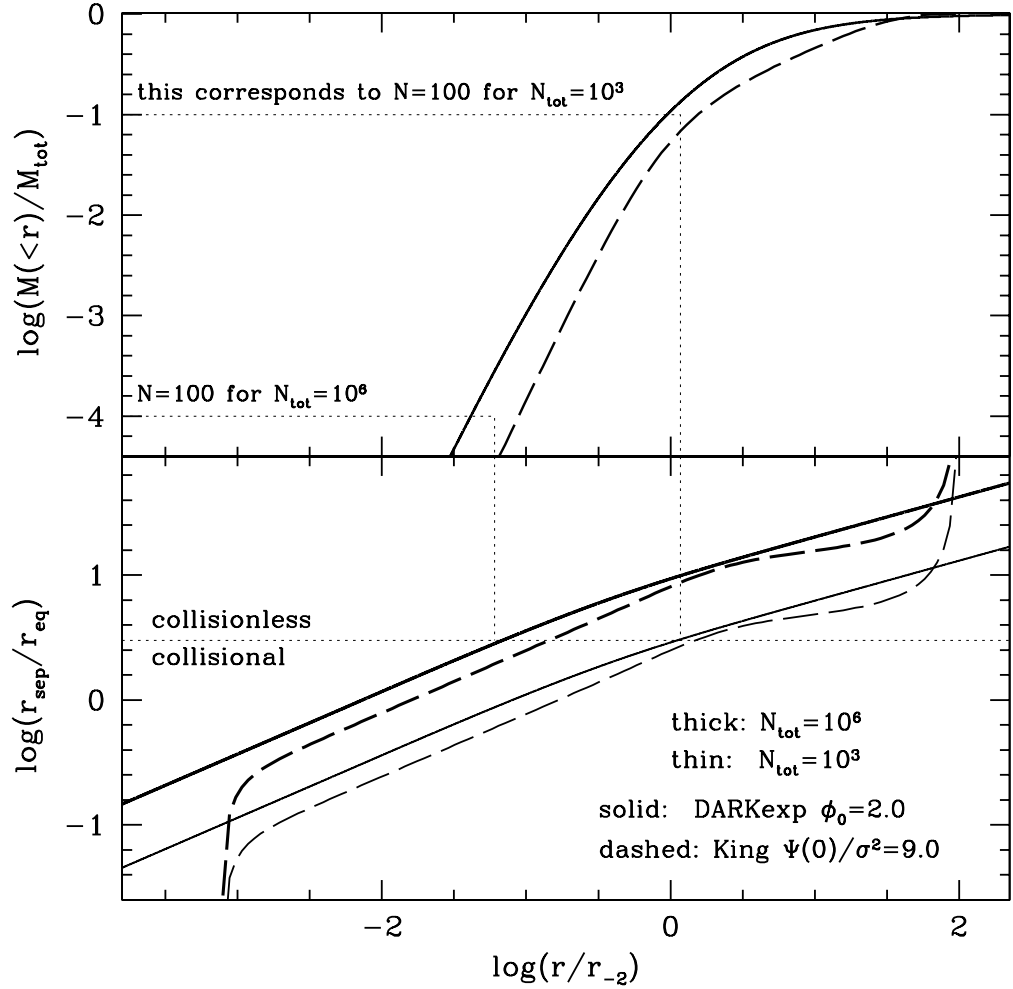}
       \caption{A comparison of two order of magnitude collisionallity tests: 
top panel is based on $t_{dyn}$ and $t_{2bd}$ time-scales, bottom panel is 
based on length-scales. The two estimates agree. See Section~\ref{coll}
for details.}
     \label{collKD}
  \end{center}
\end{figure*}

It is interesting to see how the two collisionality criteria, one
based on time-scales and the other based on length-scales compare. In
Figure~\ref{collKD} the top panels show the results of the time-scale
argument, and the bottom panel, the length-scale argument. The top
panel plots the fraction of the enclosed mass, or the fraction of the
total number of particles as a function of radius. The horizontal axis is in 
units of $r_{-2}$, the radius at which the logarithmic space density slope
$\gamma=-d\log(\rho)/d\log(r)$, is equal to 2. Above, 
it was determined that central regions where $N<100$ will be
largely collisional. The horizontal dashed lines indicate this level
for representative $N_{tot}$ values of $10^3$ and $10^6$. The bottom
panel plots $r_{sep}/r_{eq}$ vs. radius. Using the length-scale
argument above, it was determined that $r_{sep}\approx r_{eq}$ divides
collisional from collisionless regimes.  In the bottom panel, that
division is represented by the dashed horizontal line at
$r_{sep}=3r_{eq}$.  This specific relationship between $r_{sep}$ and
$r_{eq}$ guarantees agreement between the time-scale and length-scale
viewpoints for $N_{tot}=10^3$ and $10^6$ systems, and for DARKexp and
King density profiles.  The vertical dashed lines highlight that in both 
the top and bottom panels the collisional/collisionless transition occurs 
at the same location in the system, for a given size and model.

We note that instead of $t_{2bd}$ as an estimate of the two-body relaxation 
time-scale one could have chosen the Spitzer mean time-scale \citep{s87} 
which is based on the diffusion rate of particles through phase-space due 
to encounters,
\begin{equation}
t_{sp}=0.34\,\frac{\sigma^3}{G^2\,m\,\rho\,\ln\Lambda}.
\end{equation}
Here, $\sigma$, $m$, $\rho$ and $\ln\Lambda$ are the velocity dispersion,
average stellar mass, mass density, and the Coulomb logarithm, respectively.
For the DARKexp and King density profiles of a range of shape parameters,
$t_{sp}$ is larger than $t_{2bd}$ by at least a factor of $10$ over the
relevant radial ranges. Using the Spitzer time-scale would not have produced
an agreement between time- and length-scales that we have in Figure~\ref{collKD},
but since $t_{sp}$ is always longer than $t_{2bd}$ it makes two-body effects
even less important than we have estimated.

We conclude that both time- and length-scale considerations lead to the
same assessment of the collisionality of a system or part thereof, if 
$t_{2bd}$ is used. And, regardless of how two-body relaxation is 
quantified, both approaches suggest that in a typical 
cluster with $N_{tot}>10^4$, most of the body of the cluster, with the 
possible exception of the very center, would evolve collisionlessly.

\section{Models}\label{mo}

\subsection{King-Madsen models}\label{km}

Even though King models are physically motivated and have a full
dynamical description provided by the King distribution function (DF),
there did not exist a rigorous derivation of that DF until \cite{m96}.
Madsen, like King and others before him, assumes that repeated
two-body encounters in GCs require that one start with the Maxwellian 
velocity distribution and hence Maxwell-Boltzmann statistics which deal 
with classical particles that do not obey the exclusion principle.

Madsen finds that the equilibrium state can be obtained as the maximum
entropy state.  Unlike the standard derivation, which assumes that the
occupation numbers in all energy states are always large and hence
Stirling approximation is valid, he argues that no approximation 
should be made.  Following \cite{s94}, the exact form of the resulting
DF is derived, $f(E)=[Ag~\exp(-\beta E)]$, where $[.]$ means round
down to the nearest integer, and $A$, $g$ are constants, 
and $\beta$ is the constant inverse temperature of the system.
This DF has discrete `steps' that are especially pronounced at low
occupation numbers. Its smooth version \citep{hw10,bw12} is
indistinguishable from the King DF.  \cite{m96} shows that when
combined with the Poisson equation, his DF gives density profiles that
are very similar to King's, and \cite{hw10,bw12} show that the smooth
version of Madsen DF results in density profiles that are virtually
identical to the standard \cite{k66} profiles. 

Note that in the case of the King DF, `$-1$' was added by hand to 
imitate the action of the tidal field.  In Madsen's derivation, the
corresponding modification of the DF results from a proper treatment 
of low occupation numbers.

\subsection{DARKexp models}\label{De}

It has long been realized that the smooth appearance of elliptical galaxies,
which consist of tens of billions of stars, cannot be due to two-body relaxation
processes, like the rare strong encounters between stars, or the numerous
weak encounters. Some faster acting relaxation mechanism had to be at work
to ensure that the bulk of the ellipticals relax in well under a Hubble time.
This mechanism would involve a rapidly changing potential of the system, which 
would induce individual stars to change energy. Because the mechanism has to 
act fast, it was called `violent' relaxation by \cite{lb67}. It is an example 
of collisionless relaxation, because particles exchange energy with the global 
potential and not through strong or weak `collisions' among themselves. 

If we care only about the final state of the system and not the
full dynamical history, we can use the tools of
statistical mechanics. This approach to collisionless self-gravitating
systems was originally put forward by \cite{og57}, and \cite{lb67},
who argued that the final steady-state would correspond to the most
likely state. \cite{lb67} incorporated the collisionless nature of the
stellar flow through an exclusion principle in phase-space: the
collisionless Boltzmann equation states that a collisionless fluid is
incompressible, hence phase-space elements cannot be superimposed.
In the non-degenerate limit, Lynden-Bell's theory predicted that the
maximum entropy states would be isothermal spheres, as was also found
by \cite{og57}. However, this result is unsatisfactory. While the
maximization procedure imposed constraints of finite mass and energy,
isothermal spheres are infinite. Furthermore, elliptical galaxies
look nothing like isothermal spheres.  Several ways out of this
problem were proposed in the following decades.  (See the \cite{hw10}
for a further discussion of the problems with the Lynden-Bell's
approach and proposed solutions.)

\cite{hw10} argue that to apply statistical mechanics to collisionless
self-gravitating systems requires one to make two important
modification to the \cite{lb67} approach. First, the proper state
space for collisionless systems, like dark matter halos and stars in
elliptical galaxies, is energy space and not the standard phase-space.
The rationale being that in a collisionless system in equilibrium the
particles' energies are fixed, so using energy space automatically
ensures collisionlessness. Second, the expression for the possible
number of states, or entropy, involves $n!$, where $n$ is the
occupation number in energy space. When extremizing entropy, it is
customary to simplify $\ln\,n!$ using the Stirling approximation,
which is valid only for large $n$. \cite{hw10} replace the Stirling
approximation with an expression that treats all $n$ regimes very
accurately; the implication being that low $n$ regime, where $n$ is 0,
1, 2, ...  is relevant in self-gravitating systems.
With these two
modifications they derive the most likely distribution in energy for
the equilibrium systems, \begin{equation} N(E)\propto
\exp(-\beta[E-\Phi_0])-1=\exp(\phi_0-\epsilon)-1.  \end{equation}
DARKexp is a single parameter family of models, with $\phi_0$ acting
as a dimensionless potential depth.  Density profiles based on DARKexp
models can be found in \cite{wh10}. Unlike isothermal spheres, DARKexp 
systems have finite mass and energy. A comparison
with the results of collisionless dark matter simulations are
presented in \cite{whw10}. The simulations are a very good match to
DARKexp models with $\phi_0$ around 4-5. 

We emphasize that both collisional and collisionless systems, as
described by the Madsen and DARKexp models, can be derived as the most
likely statistical states.  Aside from this, Madsen and DARKexp models
have one other important feature in common: their derivation requires
that the low occupation number regime is treated properly.  The reason
why this is the case for self-gravitating systems, while the Stirling
approximation is adequate for other physical systems, is not yet
definitively decided.  What we can say is that realistic
gravitationally bound systems have DFs that are truncated, leading to
regions of state space that are very sparsely populated.  It is
therefore not surprising that entropy calculations, which depend on
counting available energy states, require those sparsely populated
regions to be accurately accounted for.

\subsection{Our goals}

The primary goal of this paper is to find out whether GCs are better
fit by King or DARKexp density profiles. These are both ``first
principles'' models and not empirical fitting functions like the 
Wilson and power-law models.  If a first-principles model is shown to 
fit the mass distribution of GCs, the most straightforward conclusion
is that the physics that went into making the model applies to GCs. 
However, it is also possible that the goodness of fits is purely
fortuitous, i.e. a complex combination of diverse dynamical effects 
happen to make the model a good fitting function.

Our secondary goal concerns only those clusters that are well fit by
the King model.  Recall that the King and Madsen DFs differ somewhat
because one is continuous and the other is discrete. It is not {\it
apriori} obvious which one is appropriate for physical systems.  The
resulting King and Madsen density profiles are very similar, but not
identical, so it makes sense to ask which of the two is preferred by
real self-gravitating collisional systems. 

\section{Data}

\cite{ng06} present the largest homogeneous set of non-parametrically
estimated surface brightness (SB) profiles of 38 Galactic globular
clusters. The inner regions of the profiles are derived from archival
HST WFPC2 images, while the outer profiles are obtained from
ground-based observations. Estimating an unbiased smooth SB profile of
the inner portions of GCs is not trivial, because a small number of
very bright giant and horizontal branch stars introduces a
considerable amount of shot noise.  NG06 performed several simulations
to determine the optimal way of overcoming shot noise, as well as
photon noise. After extensive testing on synthetic data, they conclude
that a certain combination of subtraction and masking of bright stars
works best in recovering their input profiles. The profiles of the
outer radial regions were taken directly from \cite{t95}, who used
Chebychev polynomial fits to the photometric points of ground-based
data. The analysis of NG06 combines the HST and ground-based data into
continuous composite profiles of 38 clusters, which range from $2.36$
to $3.68$ decades in radius.  For each cluster they publish 100 data
points, spaced equally in $\log(r)$.

The objectives of \cite{ng06} were to estimate the central SB slope of
GCs, deproject the light of GC where possible, and estimate the
central slope of the 3D light distribution.  Hence, they derive
uncertainties on the inner slope, but do not quote errors for the full
radial range of the SB profile. 
It is probably reasonable to assume that the typical uncertainties in
SB over the whole radial range are smaller than those in the
individual HST data points, which are $\sim \pm 0.15$ mag/arcsec$^2$.
In this paper we calculate $rms$ differences between NG06 fits and our
theoretical models, and so do not use data uncertainties. Our $rms$
values span the range from $\sim 0.014$ to $\sim 0.8$ mag/arcsec$^2$
(see Table~\ref{table1}), which can be compared to the above quoted
approximate uncertainty.

\section{Fitting GC surface brightness with DARKexp and King models}\label{twoD}

Dynamical studies of GCs are consistent with them having no dark matter.  
The main source of radial $M/L$ variation in GCs is mass segregation. 
Even in the presence of mass segregation, radial $M/L$ variations need
not be large. Lane et al.\ (2009, 2010) find only small radial
variations in $M/L$ in their GC sample. In  the radial range that the authors 
consider trustworthy their estimate of $M/L$ varies by less than a factor of 
$1.5-2$, and is consistent with being constant.  Furthermore, 
the \cite{ng06} procedure of obtaining smooth SB profiles relies mostly 
on main sequence stars, and so minimizes the effects of mass segregation. 
In this work we assume that $M/L$ is constant with radius, and hence 
the shape of the SB profiles gives the shape of the radial mass distribution 
in these systems. One has to keep in mind that this is an approximation,
and the true $M/L$ must have some radial dependence.

We use the non-parametrically smoothed profiles of NG06 as our input.
We fit the SB of each Globular Cluster with 2D projected DARKexp and
King models. DARKexp models are characterized by a single shape
parameter, $\phi_0$, a dimensionless potential depth. King models are
also characterized by a single shape parameter, $\Phi(0)/\sigma^2$, a
dimensionless combination of central potential and system's constant
velocity dispersion. An alternate parameterization is through the
concentration parameter, $c_K\equiv\log(r_{tidal}/r_{core})$; there is
a monotonic relation between $\Phi(0)/\sigma^2$ and $c_K$ for King
models. 

DARKexp models have a simple analytical expression for the energy
distribution $N(E)$, and King models have a simple expression for the
distribution function, $f(E)$. However, neither model has an analytical 
expression for the corresponding density profiles, so these have to be 
obtained numerically, even for isotropic systems. We first calculate a 
library of DARKexp and King models and then compare them to the input 
profiles from NG06.

Our fitting is done in the space of log(SB) vs. log$(r)$, not the
corresponding linear quantities, and the $rms$ deviations between GCs
and models are also calculated in the log space.  While DARKexp and
King models are one parameter families, we fit for three parameters,
one shape parameter, and radial and SB normalizations. We use the full
radial range for all 38 clusters presented in NG06, and give each of
their 100 points equal weight. 

Table~\ref{table1} presents our results. The core collapse status of
the cluster is denoted by the labels c (collapsed) and c? (possibly
collapsed), and were taken from NG06. In general, a cluster is 
classified as core collapsed if it has a cuspy, or steep SB profile 
extending to the center \citep{mh97}. In this paper we 
label clusters c and c?, but make no judgement as to their physical
state. The second and third sets of
columns, separated by vertical lines, show the DARKexp and King model
fit parameters. DARKexp $\phi_0$ and King $\Phi(0)/\sigma^2$ and $c_K$
are described above. In order to compare the two models directly, we
have defined a new concentration parameter, $c_{D13}$ and $c_{K13}$ as
~~$c_{13}=\log[r(\gamma\!=\!3)/r(\gamma\!=\!1)]$.
These concentration values are
also shown in the Table. Next to the $rms$ values for each of the two
models, we quote, in parentheses, the separate $rms$ for the inner and
the outer regions of the radial profile.  These regions each contain
50 of the total 100 points presented in NG06. These 
$rms$ values are evidence that neither the inner nor the outer portion 
of the fit dominates the total $rms$.

\begin{figure*}
  \begin{center}
    \leavevmode
      \epsfxsize=15cm\epsfbox{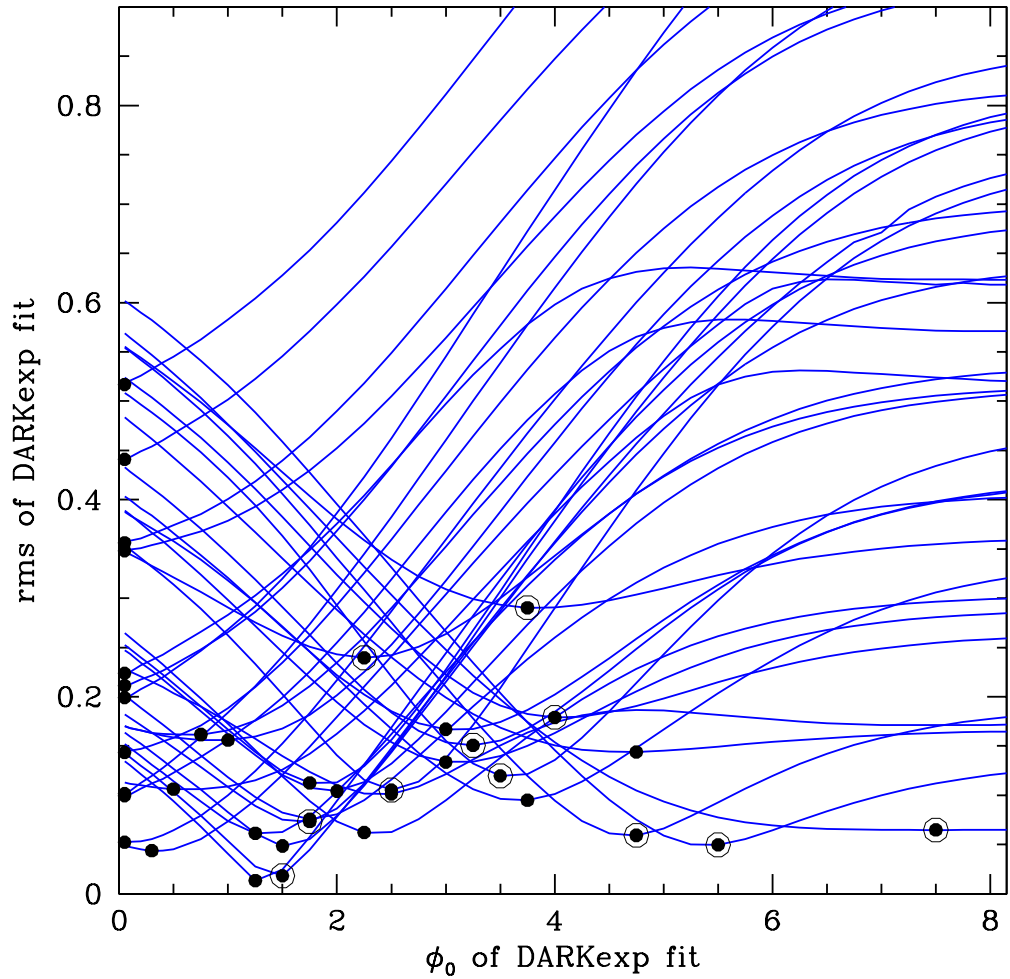}
\caption{The curves represent DARKexp fits to 38 GCs. The best-fit DARKexp model for 
each cluster is marked by a solid dot. Eleven clusters have best-fit $\phi_0=0.05$;
these are not to be trusted because 0.05 was the smallest $\phi_0$ value considered
because the outer SB profiles of most of these clusters are steeper than what DARKexp 
can have, so no DARKexp model can provide a good fit. The circled dots represent
core collapse GC.}
\label{mcmcfitDARKexp}
  \end{center}
\end{figure*}

\begin{figure*}
  \begin{center}
    \leavevmode
      \epsfxsize=15cm\epsfbox{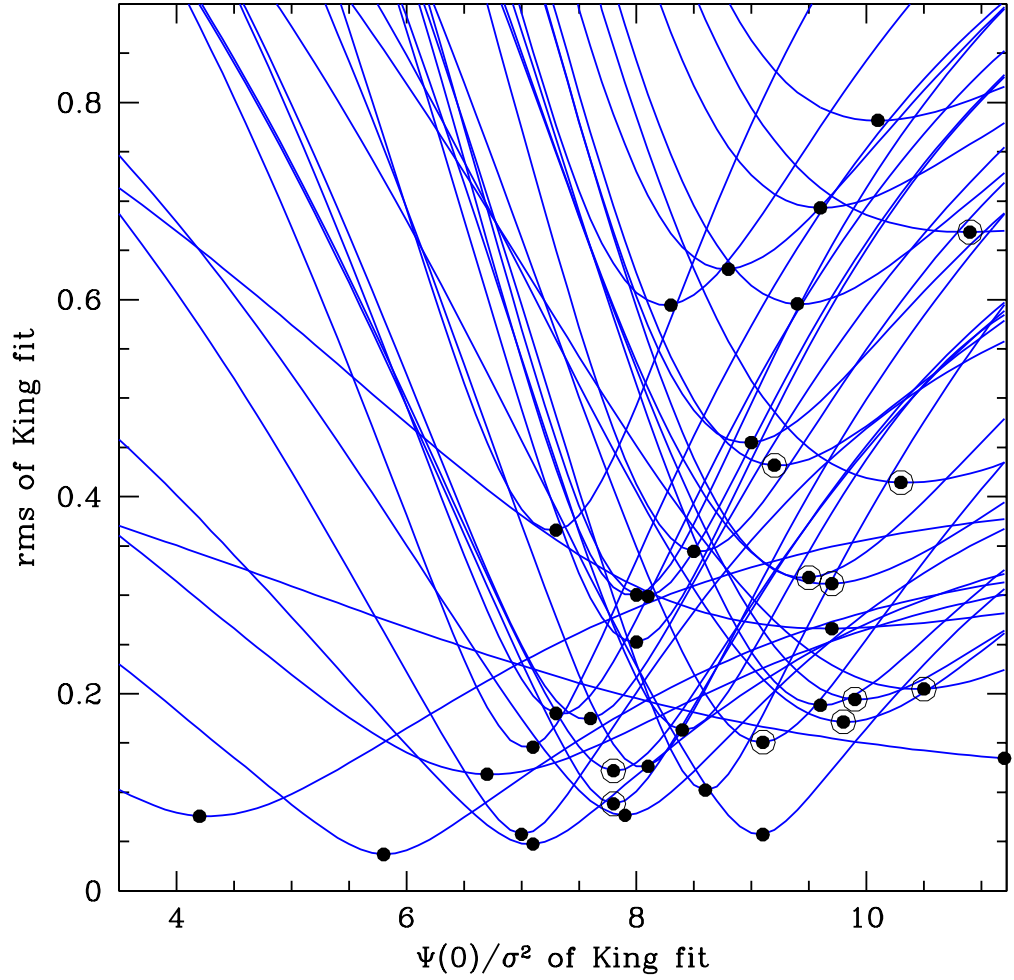}
\caption{Similar to Figure~\ref{mcmcfitDARKexp}, but for King fits. The cluster with
best-fit $\Psi(0)/sigma^2$ beyond 11 (at the right edge of the plot) is NGC 6352;
Figure~\ref{fp06}. It has an upturn in SB at large radii, which is difficult to
fit even with very large concentration King models.}
\label{mcmcfitKing}
  \end{center}
\end{figure*}

\begin{figure*}
  \begin{center}
    \leavevmode
      \epsfxsize=15cm\epsfbox{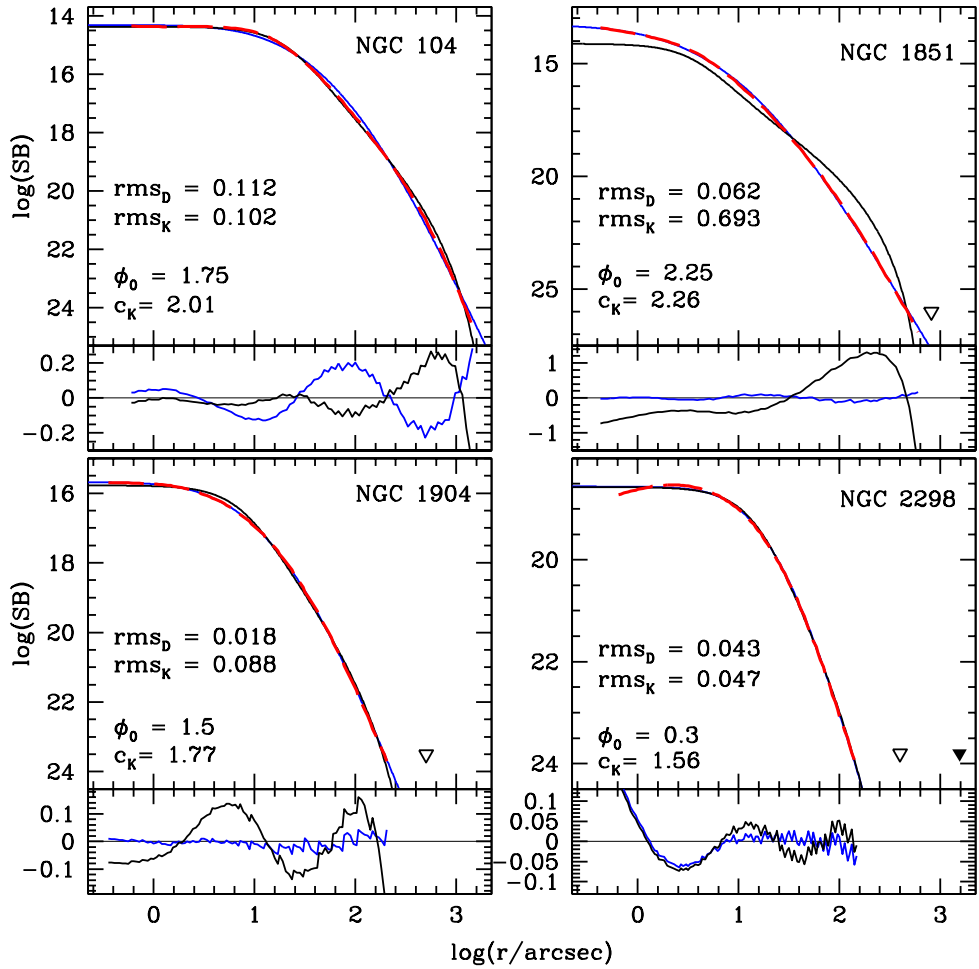}
\caption{DARKexp (blue) and King (black) best-fits to four GCs. The red dashed curve is
the Noyola \& Gebhardt (2006) data, which consists of HST and ground-based observations, jointed and smoothed
by those authors. The surface brightness (SB) is in mag/arcsec$^2$ in the $V$ band. The $rms$ 
between the data and models are indicated in the plot, together with the values of the 
dimensionless potential depth for DARKexp model, $\phi_0$, and the King concentration 
parameter, $c_{K}$. The fit residuals are shown in the bottom insets. The vertical scale 
is the difference in SB, in mag/arcsec$^2$; the its span varies between panels.
The empty downward triangles indicate the King model truncation radii taken from Gnedin \& Ostriker(1997).
The filled triangles indicate the truncation radius estimated from the strength of the 
Galactic tidal field; see Section~\ref{afew}.}
\label{fp01}
  \end{center}
\end{figure*}

\begin{figure*}
  \begin{center}
    \leavevmode
      \epsfxsize=15cm\epsfbox{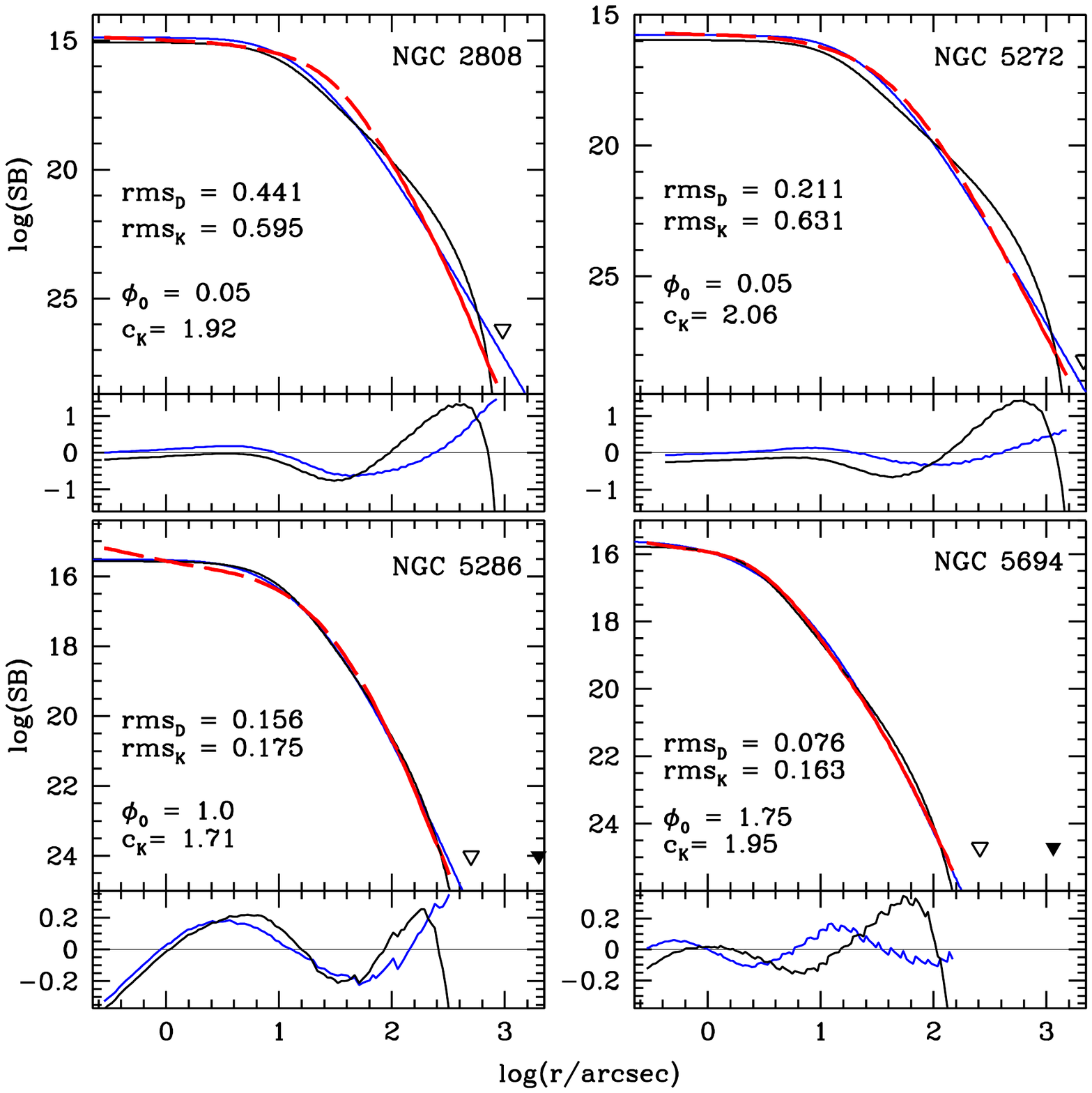}
\caption{Similar to Figure~\ref{fp01}.}
\label{fp02}
  \end{center}
\end{figure*}

\begin{figure*}
  \begin{center}
    \leavevmode
      \epsfxsize=15cm\epsfbox{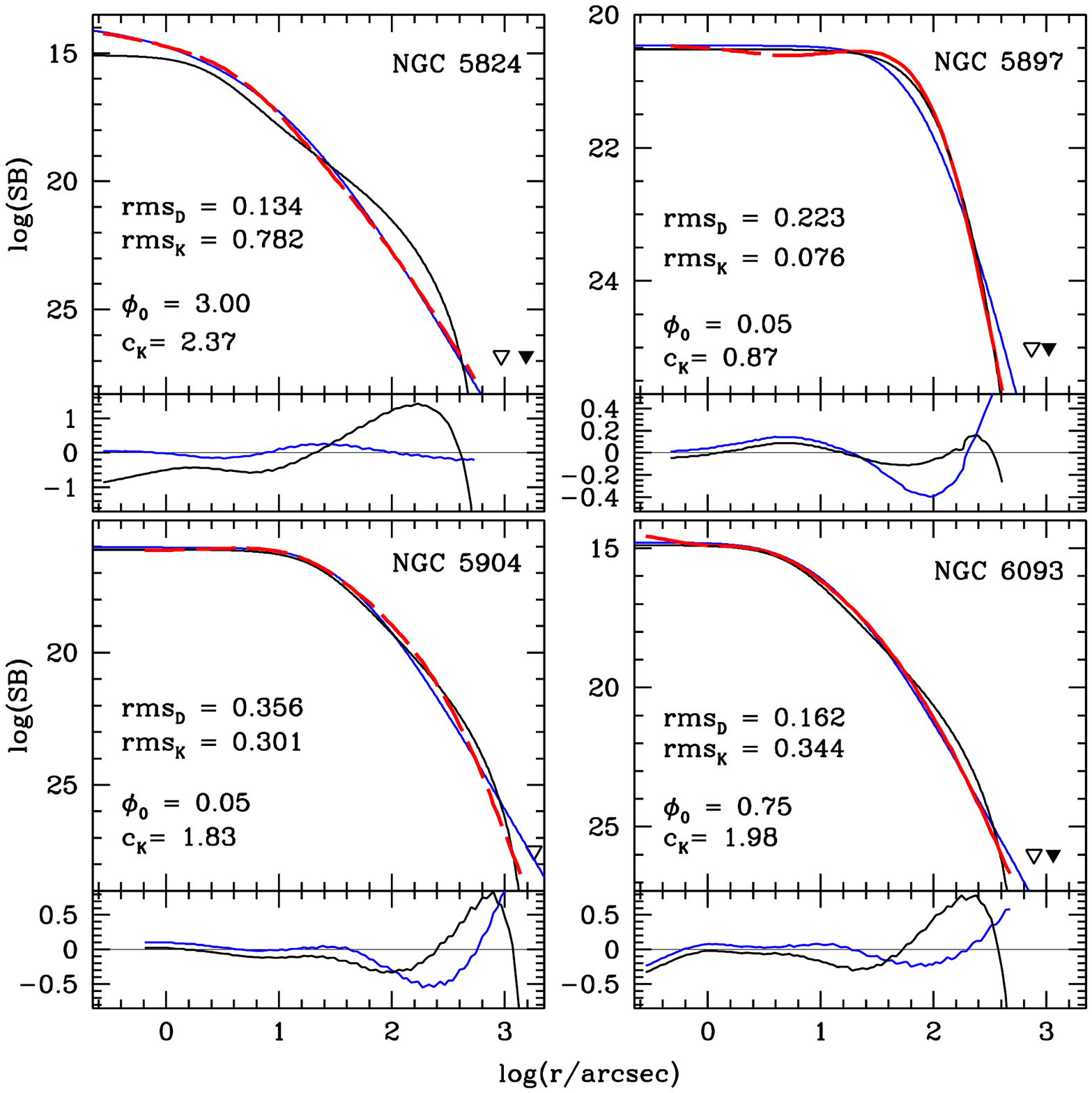}
\caption{Similar to Figure~\ref{fp01}.}
\label{fp03}
  \end{center}
\end{figure*}

\begin{figure*}
  \begin{center}
    \leavevmode
      \epsfxsize=15cm\epsfbox{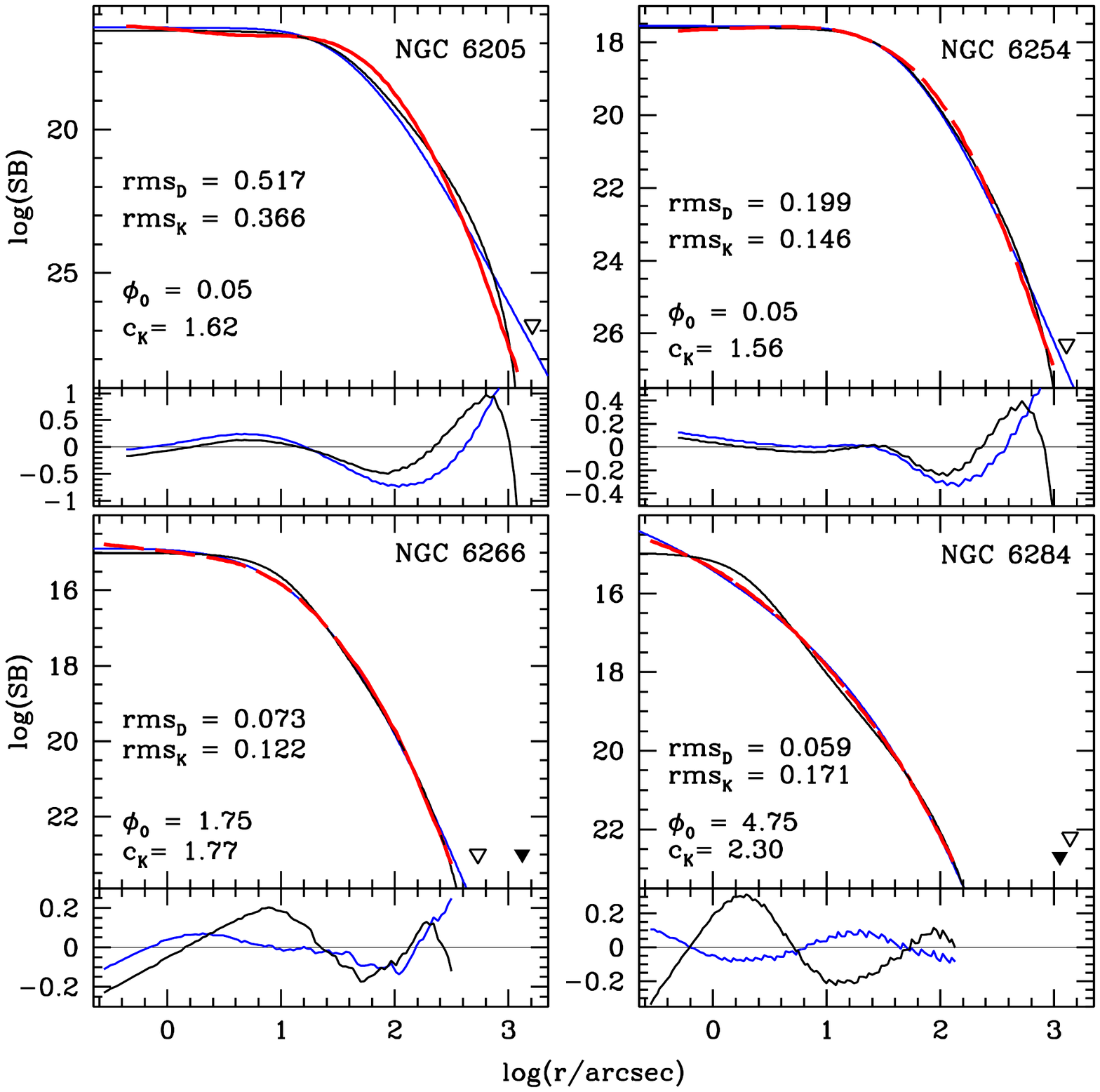}
\caption{Similar to Figure~\ref{fp01}.}
\label{fp04}
  \end{center}
\end{figure*}

\begin{figure*}
  \begin{center}
    \leavevmode
      \epsfxsize=15cm\epsfbox{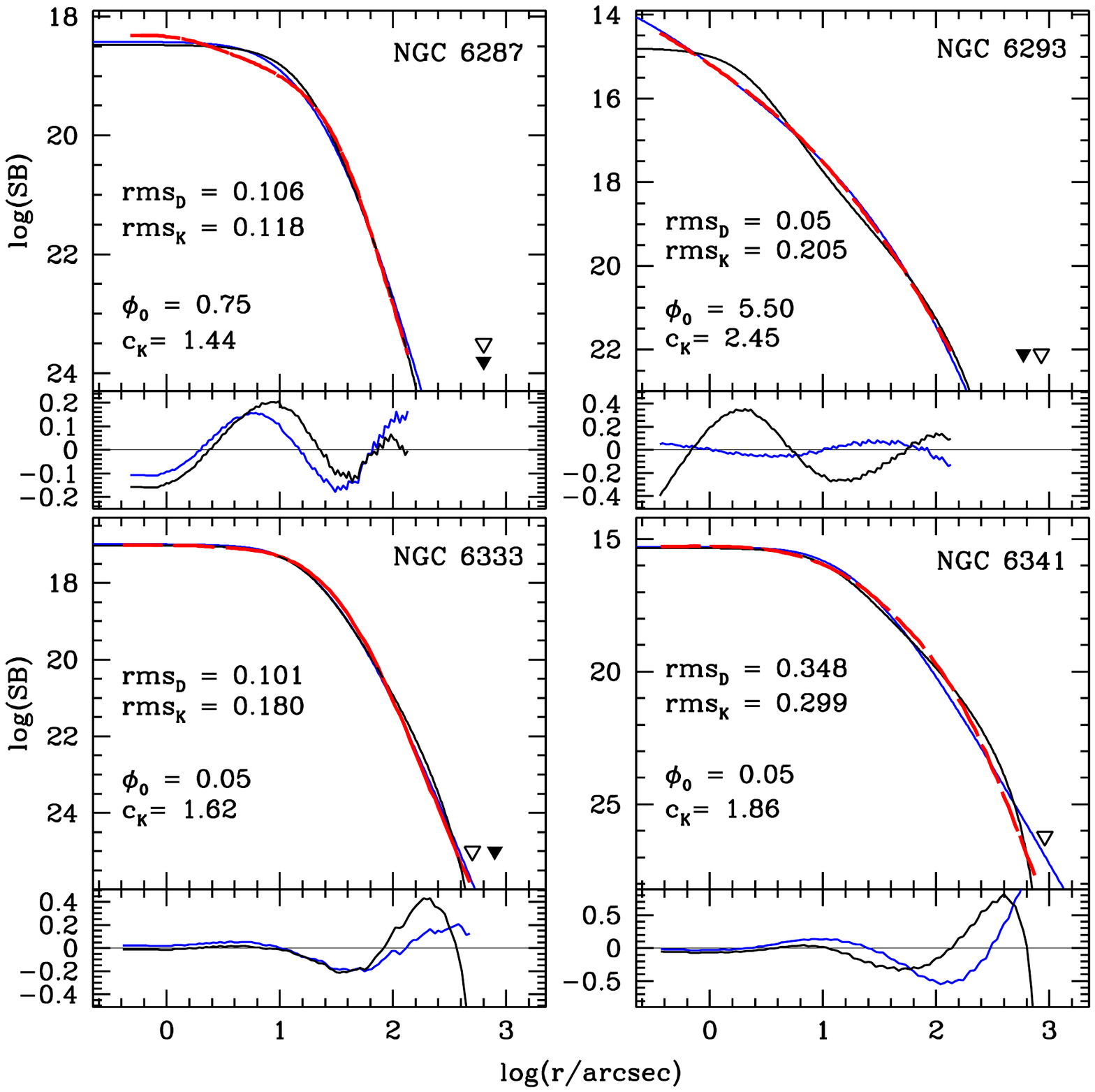}
\caption{Similar to Figure~\ref{fp01}.}
\label{fp05}
  \end{center}
\end{figure*}

Figures~\ref{mcmcfitDARKexp} and \ref{mcmcfitKing} summarize our fits
for DARKexp and King models, respectively.  The horizontal axis is the
potential depth (shape parameter) of the model, and the vertical axis
is the $rms$ of the fit. The best-fit model for each of the 38
clusters is represented by a solid dot, which is circled if the
cluster is considered to be core collapsed.  The best fits for
individual clusters are shown in Figures~\ref{fp01}-\ref{fp10}. Red
dashed curves are the NG06 data, blue curves are best-fit DARKexp, and
black are best-fit King profiles. The fit residuals (in $\Delta$
mag/arcsec$^2$) are shown in the bottom insets of each panel. Note
that some residual curves show small amplitude fluctuations with
radius, that look like noise. These come from the NG06 surface
brightness data which is quoted to only 4 significant digits. This
limited precision shows up as `noise', which is especially visible
when the fits are good.  We do not smooth over this noise here, but do
smooth over it in Section~\ref{threeD}.

All results presented in this Section indicate that most clusters are
better fit by DARKexp than King.  Additionally, about a quarter of the
clusters are fit very well, over the entire available radial range, by
DARKexp. Figure~\ref{comprms} plots best-fit $rms$ of DARKexp and King
models; only eight clusters are better fit by King.
Note that all eleven core collapse GCs are better fit by DARKexp than 
King models. Even if these are taken out, there is still a clear 
preference for DARKexp for non-core collapse GCs. Recall that the
DARKexp model was built to reproduce the equilibrium states of
collisionless systems, like dark matter halos.  Apparently, many GCs
are also well fit by it. We interpret this result to mean that 
the overall SB profiles of GCs may be
less affected by collisional processes among its member stars
than previously assumed. 

To further compare DARKexp and King fits, in Figure~\ref{residuals} we
show residuals from best-fits for all 38 clusters, for DARKexp (top
panel) and King (bottom panel), respectively.  The horizontal axis is
normalized by the core radius of each cluster, which was defined by
NG06 as the radius where the SB drops to half its central value. Note
that this core radius definition is non-parametric and thus unrelated 
to the definition of the core radius in King models; the
half-light radius was determined from non-parametric smoothed
profiles. In addition to showing that DARKexp residuals are smaller,
the figure also suggests that the King residuals show a systematic
pattern, while DARKexp residuals are more random.  If true, then this
is an additional indication that DARKexp are a better fit, at least
typically, than King models. 

A skeptic might argue that the full radial range of a GC should not be
fit with a single profile because different dynamical processes are at
play at small and large radii. The innermost radii might be affected by 
a central black hole, or core collapse and compact binaries,
while the outermost radii may be populated by extra-tidal stars which are 
not in equilibrium with the global potential. We do not address these 
possibilities, aside from pointing out that these modifications would 
need to be described by multiple adjustable parameters.  

Before proceeding, we examine one more issue relevant for fitting, the
radial extent of the data. The goodness of fit depends on the radial
range being fit, and we have already mentioned that the shorter ranges
available in the 1970's and 1980's were well modeled by the King
profile.  In Figure~\ref{rrange}, the vertical axis shows the log of
the ratio of the radial range given in NG06, and the horizontal axis
the $rms$ of the best-fit: solid dots represent DARKexp and empty
triangles, King. The vertical solid and dashed lines in the top and
bottom halves of the plot show the average $rms$ for the DARKexp and
King fits, respectively. As expected, a larger radial range leads to
poorer fits for both models.  DARKexp is moderately better than King
for smaller radial ranges (bottom half of the plot).  As the radial
range gets larger (top half of the plot), DARKexp fits become
noticeably better fits than King.  This too points towards DARKexp
being a better descriptor of GCs.

\section{King vs. Madsen models}

Our secondary goal in this paper is to compare \cite{k66} and
\cite{m96} models.  The King and Madsen models are created by smooth
and discrete distribution functions, respectively, and show
corresponding differences in the density profiles. Since it is not
clear which of these two are more physically appropriate, we fit both
models to twelve GC that are well fit, $rms<0.15$, by King models
(except NGC 6352, see the caption of Figure~\ref{mcmcfitKing}). The
clusters that are poorly fit by King tend to be much better fit by
DARKexp (see Figure~\ref{comprms}), and so are unlikely to be better
fit by the Madsen model. 

Figure~\ref{bestfitM} presents the results. The blue solid and dashed
lines, and the solid and empty circles represent King and Madsen fits,
respectively. The best fits for the same cluster are connected by a
red straight line segment. Eleven of the twelve clusters are better
fit by the King model. Because the differences between King and Madsen
profiles are small, we might have expected a breakdown closer to
50/50. The reason is probably the very sharp drop off in density that
Madsen models have at large radii, a consequence of the discreteness
of the distribution function at low occupation numbers.
Real GCs do not show such steep drop off.

\section{DARKexp vs. King models: a closer look at some GCs}\label{afew}

In this Section we consider a few specific GCs and compare the
standard description of their SB to the one using DARKexp.

\begin{figure*}
  \begin{center}
    \leavevmode
      \epsfxsize=15cm\epsfbox{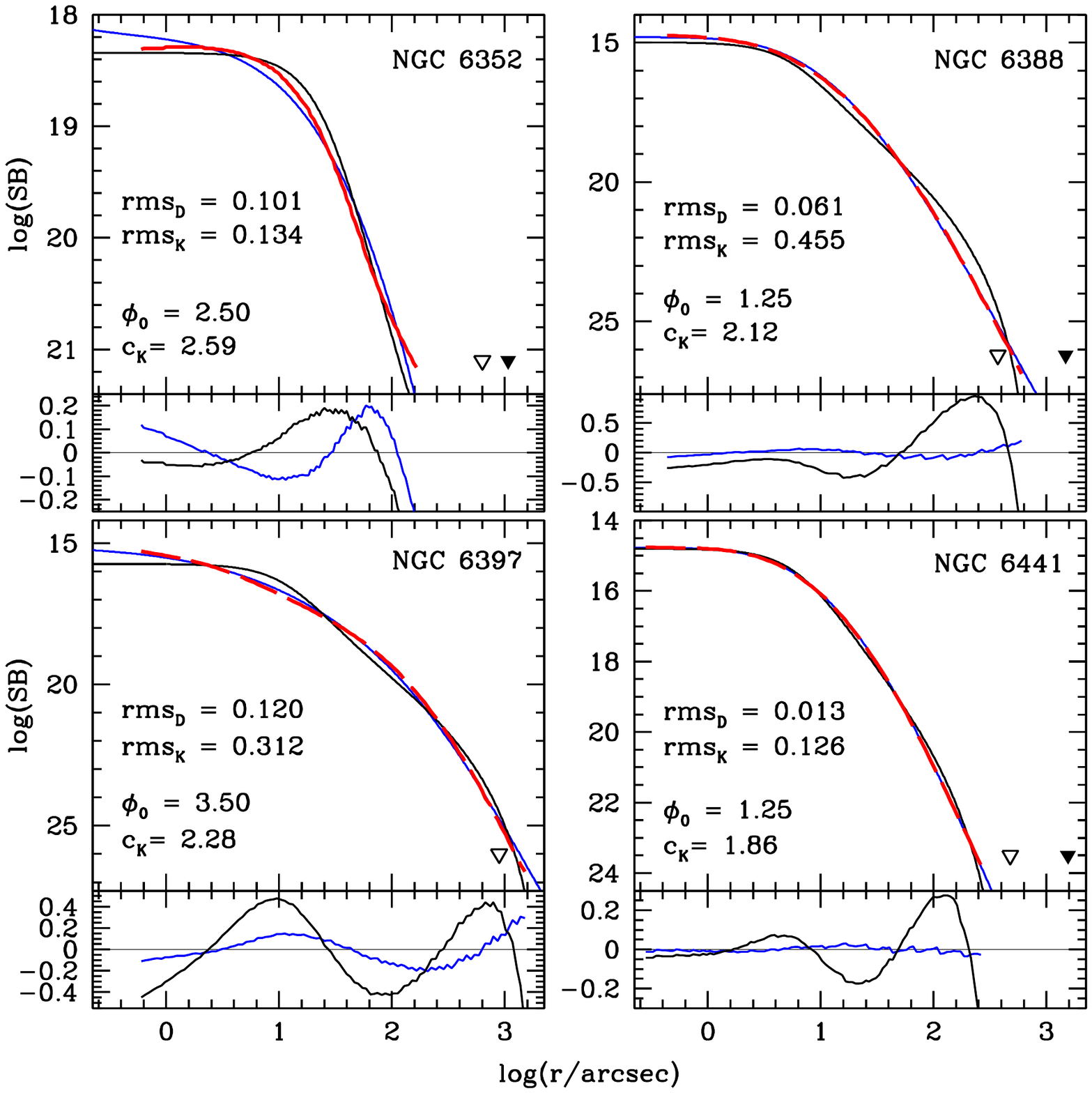}
\caption{Similar to Figure~\ref{fp01}.}
\label{fp06}
  \end{center}
\end{figure*}

\begin{figure*}
  \begin{center}
    \leavevmode
      \epsfxsize=15cm\epsfbox{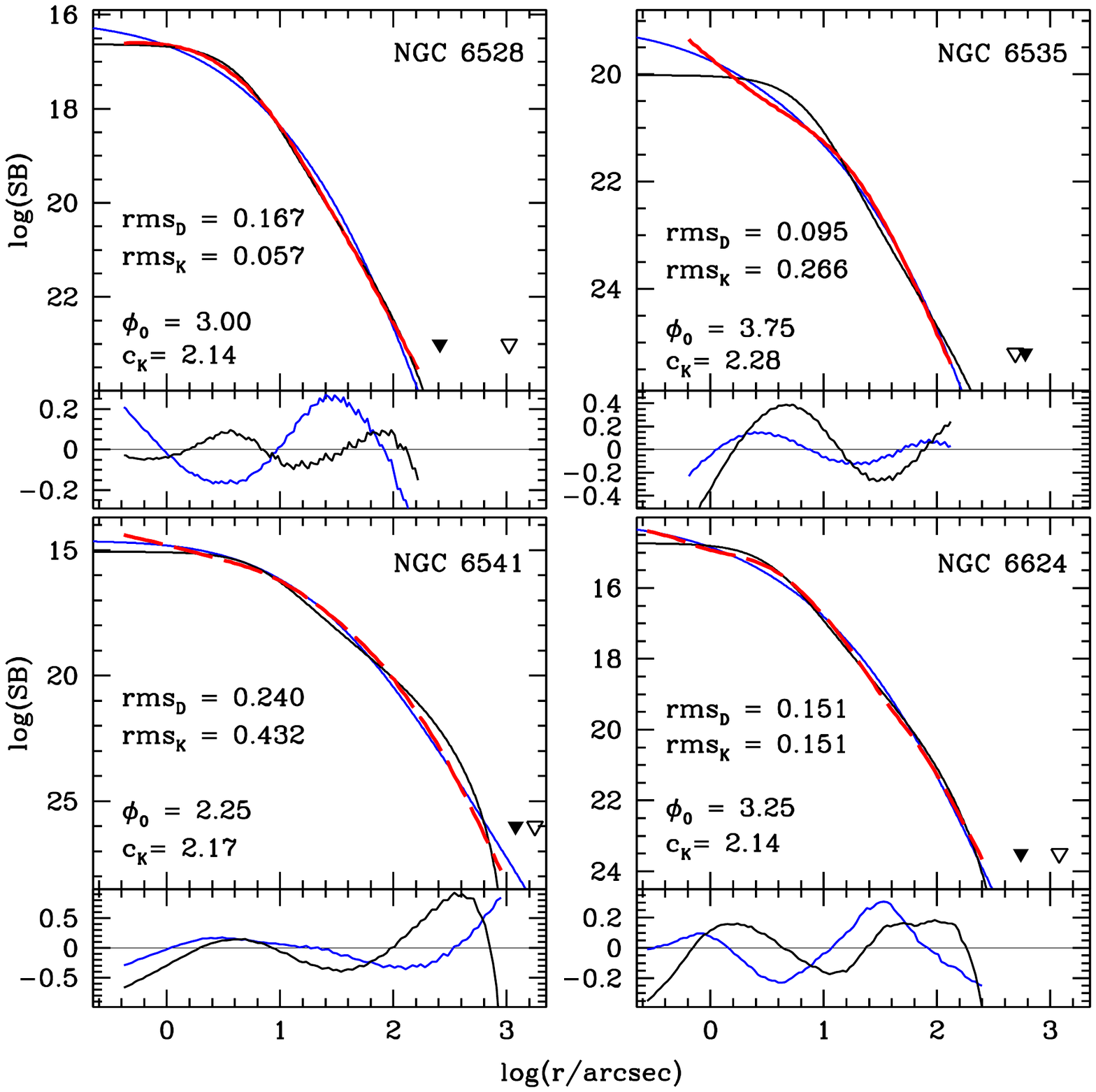}
\caption{Similar to Figure~\ref{fp01}.}
\label{fp07}
  \end{center}
\end{figure*}

\begin{figure*}
  \begin{center}
    \leavevmode
      \epsfxsize=15cm\epsfbox{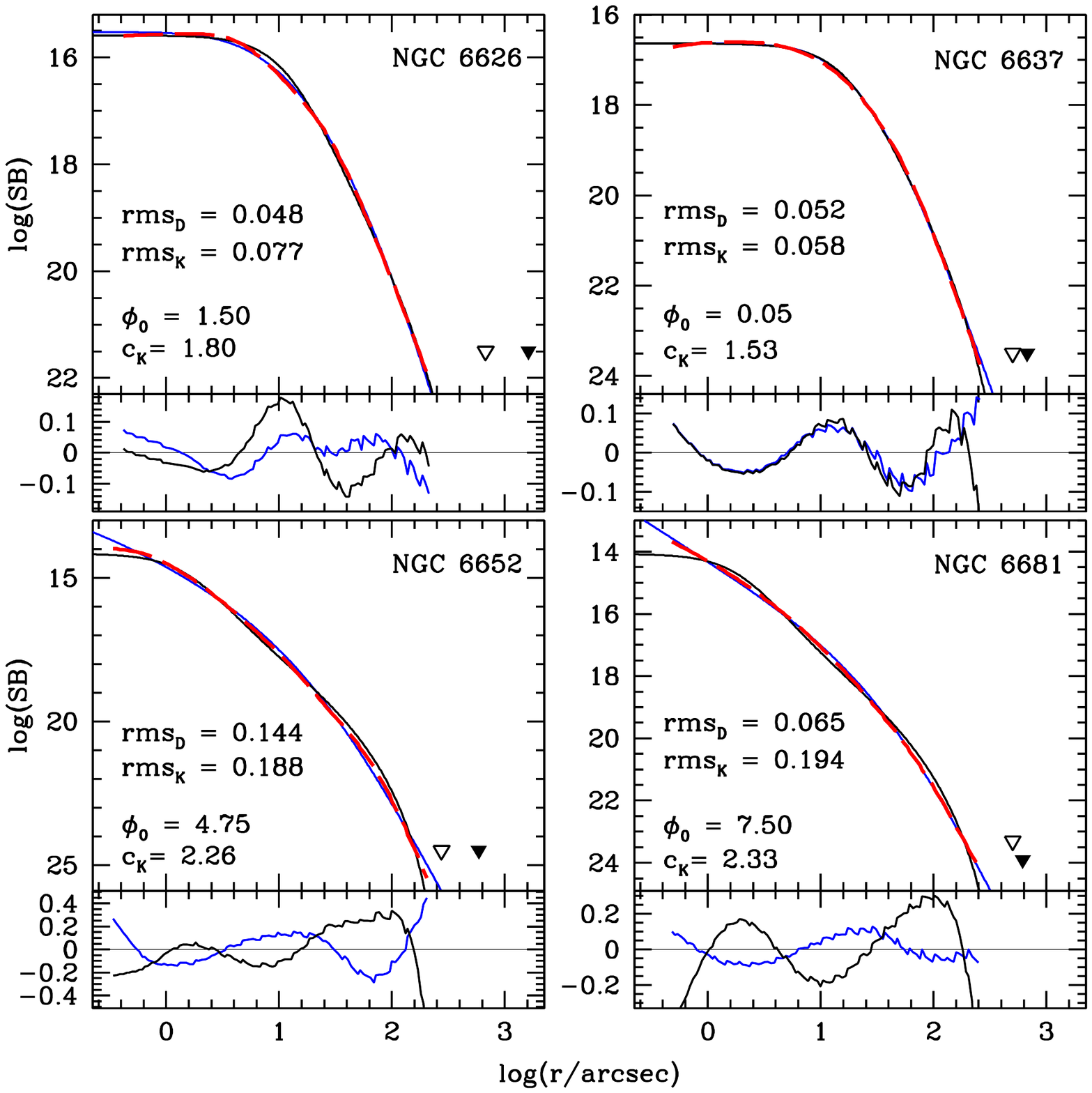}
\caption{Similar to Figure~\ref{fp01}.}
\label{fp08}
  \end{center}
\end{figure*}

\begin{figure*}
  \begin{center}
    \leavevmode
      \epsfxsize=15cm\epsfbox{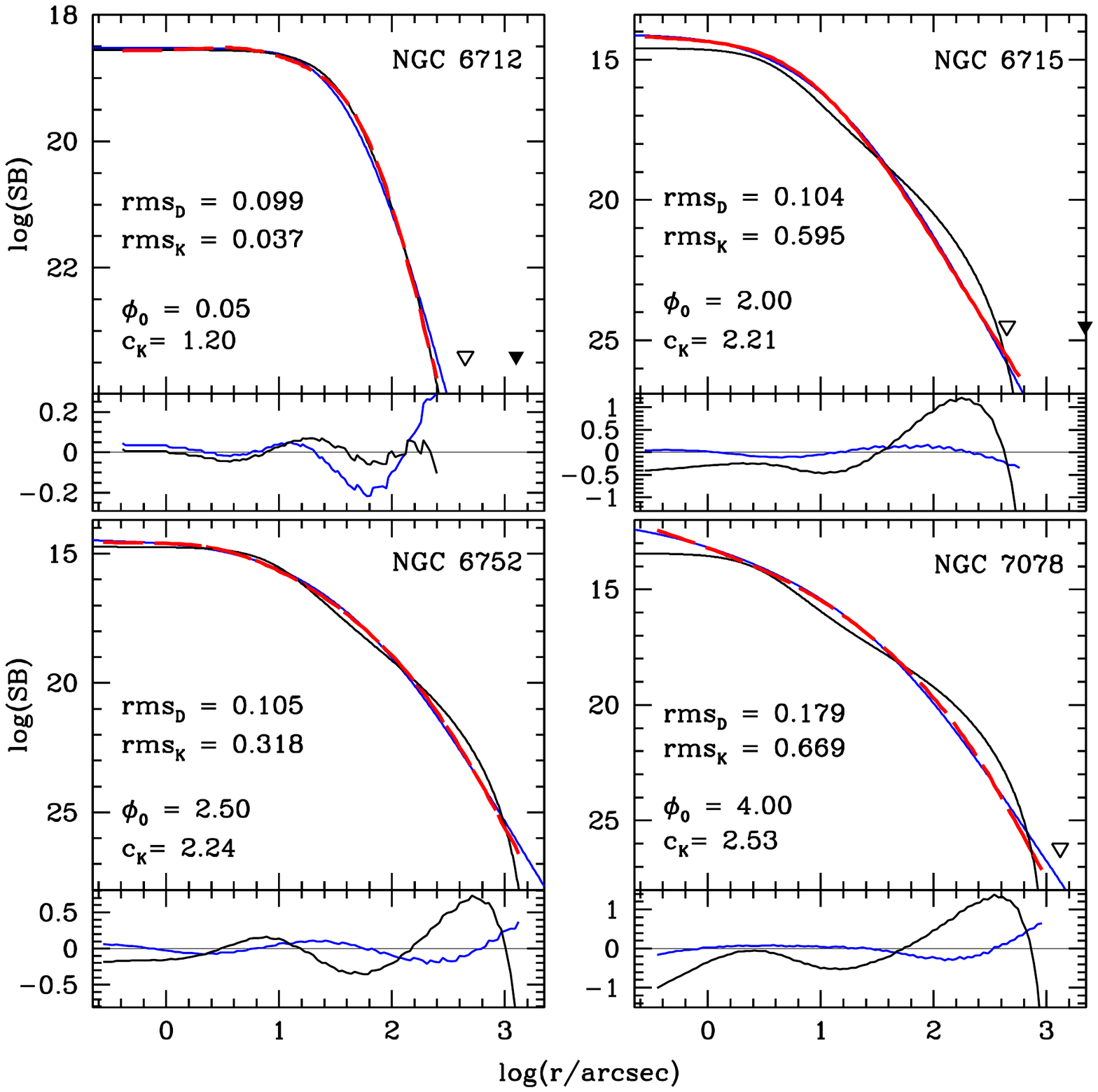}
\caption{Similar to Figure~\ref{fp01}.}
\label{fp09}
  \end{center}
\end{figure*}

\begin{figure*}
  \begin{center}
    \leavevmode
      \epsfxsize=15cm\epsfbox{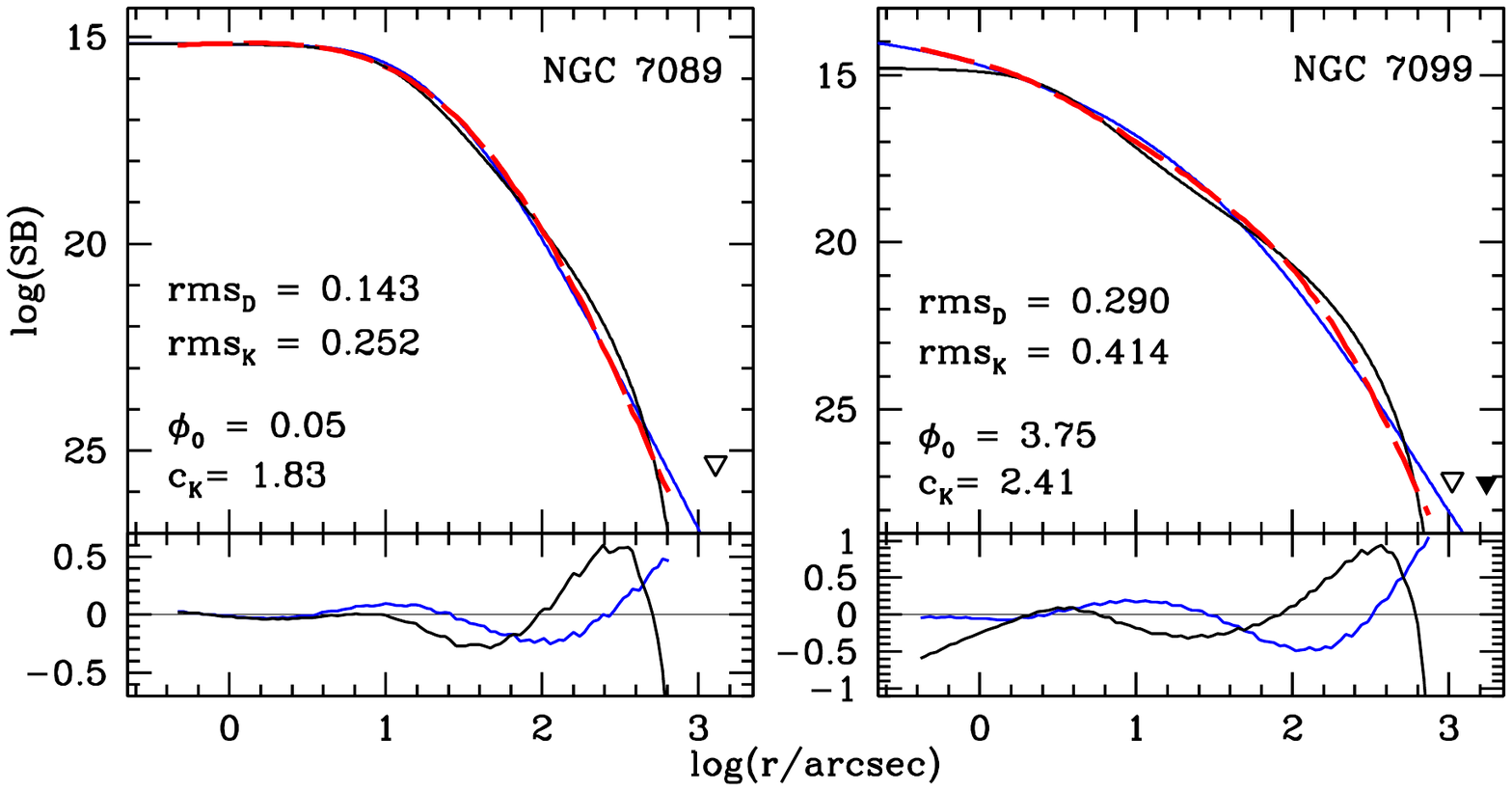}
\caption{Similar to Figure~\ref{fp01}.}
\label{fp10}
  \end{center}
\end{figure*}

\begin{figure*}
  \begin{center}
    \leavevmode
      \epsfxsize=15cm\epsfbox{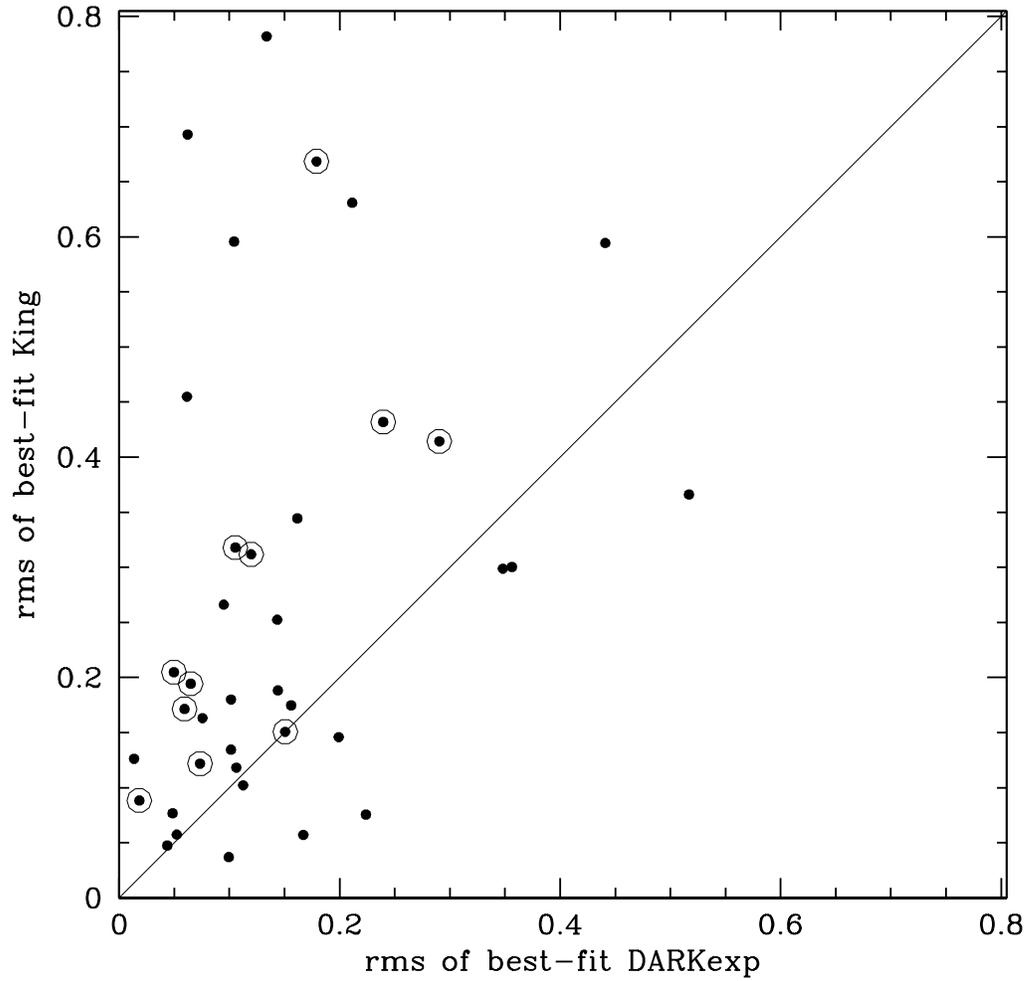}
\caption{$Rms$ of best-fit King and DARKexp models for 38 GCs. Circled dots represent 
eleven GCs that are considered to be core-collapsed. Overall, DARKexp models fit GCs 
considerably better than King models; out of 38 GCs only eight are better fit by King 
compared to DARKexp.}
\label{comprms}
  \end{center}
\end{figure*}

\begin{figure*}
  \begin{center}
    \leavevmode
      \epsfxsize=15cm\epsfbox{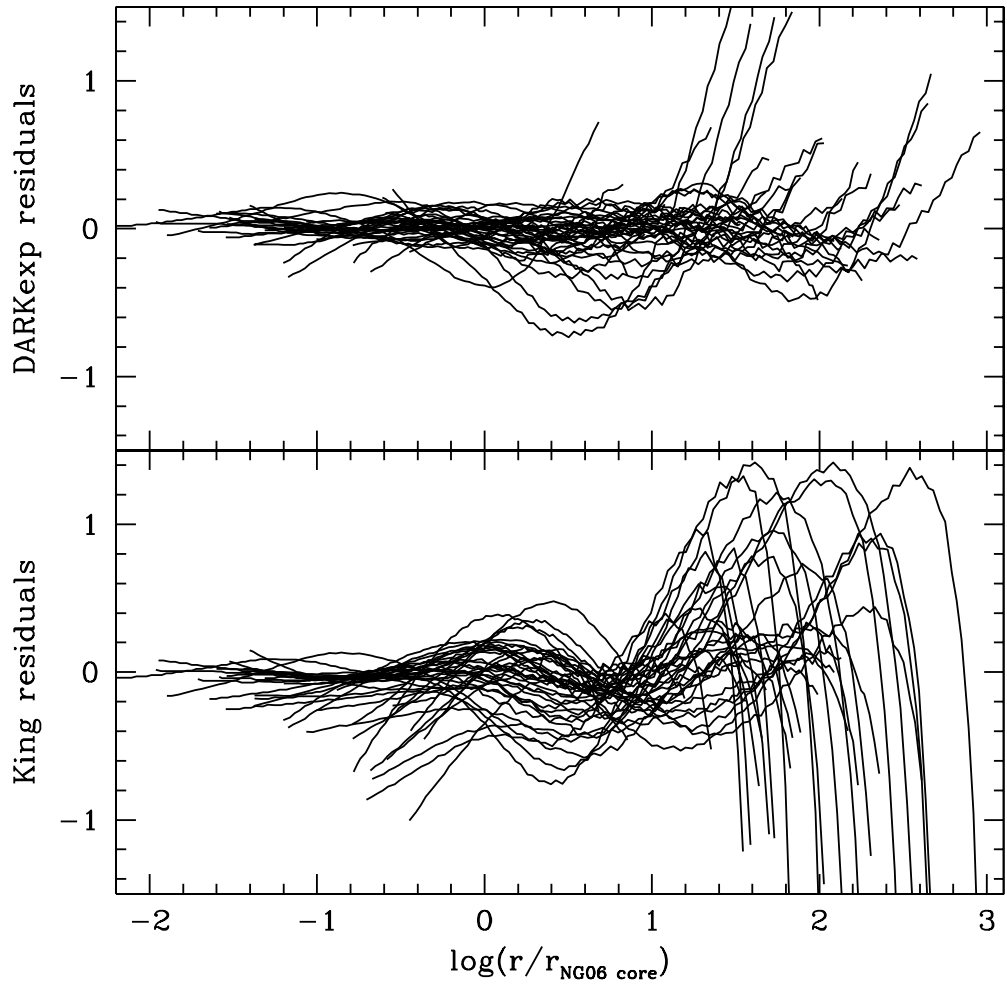}
\caption{Residuals from best-fit DARKexp and King models for 38 GCs. Radius is normalized by the 
core radius defined by NG06 as the radius where the SB drops to half its central value. This radius
is unrelated to King core radius. In addition to being larger, King residuals also show a more 
systematic pattern compared to DARKexp residuals.}
\label{residuals}
  \end{center}
\end{figure*}

\begin{figure*}
  \begin{center}
    \leavevmode
      \epsfxsize=15cm\epsfbox{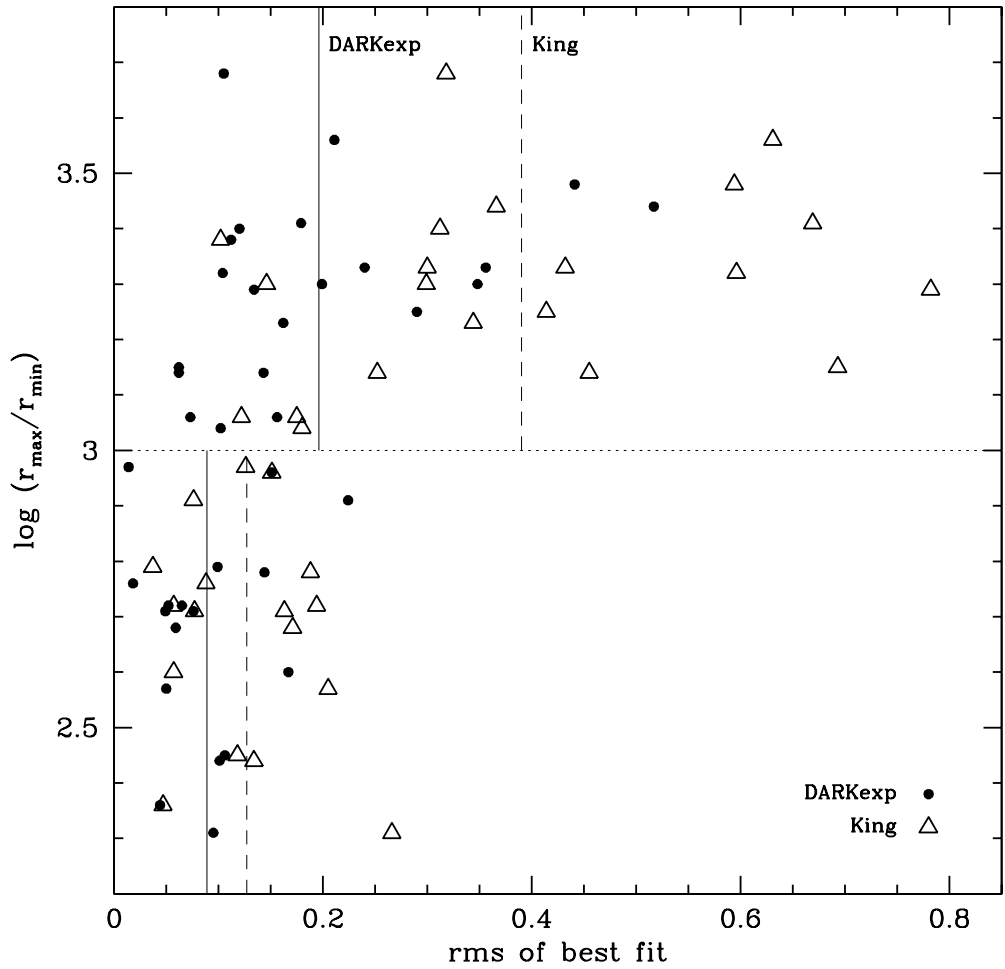}
\caption{Radial range, $\log(r_{max}/r_{min})$ of NG06 data vs. $rms$ of the best-fit for 38 GCs. 
DARKexp fits are filled dots, King fits are empty triangles. The horizontal line divides the sample 
in to two roughly equal sets. In each set the solid vertical line gives the average $rms$ for the 
DARKexp fits, while the dashed line give $rms$ for King fits.}
\label{rrange}
  \end{center}
\end{figure*}

\begin{figure*}
  \begin{center}
    \leavevmode
      \epsfxsize=15cm\epsfbox{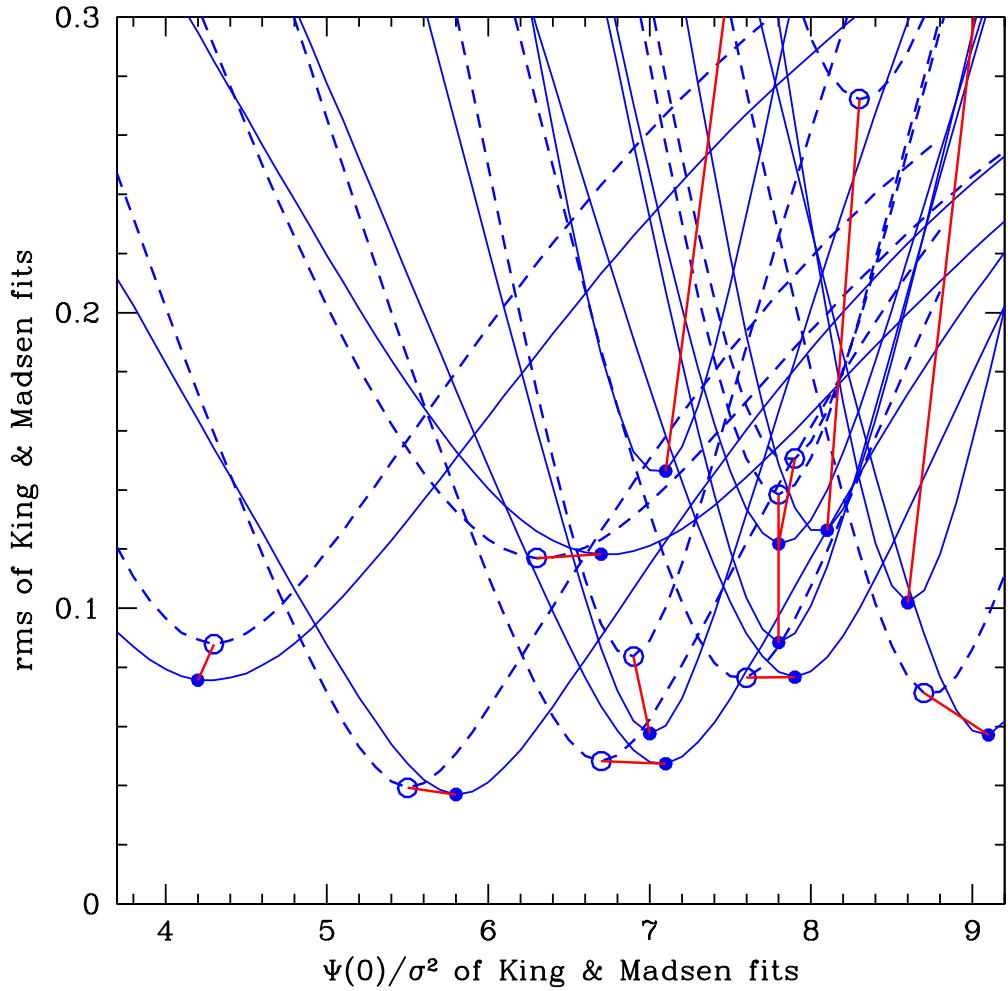}
\caption{Similar to Figure~\ref{mcmcfitKing}, but only for 12 GCs where King models are a good fit.
The solid lines and dots represent fits with King models, while the dashed curves and empty circles 
are the Madsen model fits. Red line segments connect best-fit King and best-fit Madsen models for 
the same cluster.}
\label{bestfitM}
  \end{center}
\end{figure*}

Modern data, especially HST and ground-based composite data sets such
as the one provided by \cite{ng06}, have extended radial coverage
and good resolution down to very small radii. Here we illustrate,
with one specific example, that a large radial range is often needed to
discriminate between competing models.  Consider NGC 6388, the cluster
with the largest velocity dispersion, $\sigma=18.9$ km/s \citep{pm93}
in the NG06 sample. Using data spanning $3''$ to $250''$, \citet{ii76}
fit a King model of $c_K=1.75$. Within these 2 radial decades, the King
model fits the data well.  If we truncate the NG06 data to the above
radial range, our best-fit King model has $c_K=1.77$, consistent with
earlier findings. However, the full 3.14 radial decades clearly prefer
DARKexp over King (Figure~\ref{fp06}), with $rms$ values of 0.062 and
0.455, respectively.

About 20\% of Galactic GCs show deviations from King models by having
steeper central SB profiles, {\it i.e.}, cusps instead of cores.
Moderately steep central cusps could be, in some cases, a signature of
an intermediate mass black hole (IMBH), between $10^2$ and $10^4
M_\odot$ \citep[though the presence of a shallow cusp is not a
reliable indication of the IMBH;][]{vt10}.  Steeper cusps are explained
as post-core collapse clusters. However, as NG06 point out, clusters
undergoing gravothermal oscillations should spend a small fraction of
their lives in collapsed states.  If all clusters that show cusps are
assumed to be in a collapsed state, then we are catching a
disproportionately large fraction of GC in this short-lived phase of
their evolution. Consider the GC with the steepest cusp in the NG06
sample, NGC 6681 (M 70).  The King fit has $c_K=2.3$ and $rms=0.194$.
The DARKexp fit has $\phi_0=7.5$ and $rms=0.065$.  The rather steep
central cusp is very well accommodated by DARKexp. Whether it speaks
for or against the cluster being core collapse or hosting a central
black hole is not clear, but the overall mass distribution over the
entire 2.72 decades in radius is definitely better accounted for by
DARKexp. 

Another example is NGC 6715 (M 54). \cite{i09} find density and
velocity dispersion cusps within the central 0.3pc which they
interpret as evidence of a 9400 $M_\odot$ IMBH. However, they can also
explain the cusp if the central stars have moderate radial anisotropy.
\cite{w11} find no evidence for the IMBH in the X-rays, deriving an
upper limit on the Eddington ratio  $<1.4\times 10^{-10}$. Regardless
of whether the cluster hosts an IMBH, DARKexp provides a better fit,
including the central cusp, than King; the two $rms$ are 0.104 and
0.595, respectively.


Deviations from King models at large radii are also seen.  These are
generally attributed to extra-tidal stars, since GCs are expected to
lose stars due to tidal stripping.  Several GCs do
show spectacular tidal tails. However, in some others, the
`extra-tidal' distribution is circularly symmetric, and hence at odds
with a tidally induced scenario. Such a distribution can be explained
by evaporated stars \citep{ku10,ku11}. However, a DARKexp model
also fits. For example, NGC 5694 was recently
examined by \cite{c11} using VIMOS/VLT. These authors note that they
do not see a break, or any tidal tails beyond the cluster's tidal
radius of 3.15 arcmin. Instead, the stellar distribution smoothly
continues across the tidal radius, up to 10 arcminutes from the
center. It shows an almost constant SB slope before and after the tidal
radius, which the authors fit with a power law, $R^{-3.2}$ (see their
Figure~3). In 3D this would be $r^{-4.2}$, completely consistent with
DARKexp, which has an asymptotic outer slope of $-4$, independent of
$\phi_0$. Our best-fit DARKexp model for NGC 5694 has $\phi_0=1.75$,
with $rms=0.076$, compared to King model $rms=0.163$. 

That truncation radii based on King model fitting may not correspond
to tidal truncation due to the Galaxy is generally acknowledged.
In Figures~\ref{fp01}-\ref{fp10} the empty downward triangles indicate
the King model truncation radii taken from \cite{go97}. The filled
triangles indicate the truncation radius estimated from the strength
of the Galactic tidal field, $r_{t}^3=r^3 (M_{GC}/[3M(<r)])$,
where $r$ is the Galactocentric distance, $M_{GC}$ is the GC's mass
obtained from its absolute luminosity in $V$ and a constant mass-to-light
ratio of $M/L=2.5$ in solar units, and $M(<r)$ is the mass of the Galaxy 
interior to the current (i.e. not peri-center) location of the cluster. 
(Note that some truncation radii lie outside the limit of the figures.)

Some GCs that are not well fit with King models and show signs of mass
segregation are fit with multi-mass King models; see for example,
\cite{df76}. These more complex models are especially needed for
clusters that cover a larger dynamic range in SB, up to 5 decades. NGC
5272 (M3) is one example.  \cite{df76} show that the inner radial
range can be fit with $c=1.29$ King, while the outer with $c=1.98$.
Their actual fitted model has 10 mass ranges, for the data that span
2.4 decades in radius.  NG06 data spans 3.6 decades, and the DARKexp
fit has $rms=0.211$, which appears comparable to the residuals of the
10-mass King profile (see Fig 3 of \cite{df76}).

An an illustration of how different estimates of global parameters can be depending
on the model fitted, we quote effective radii and enclosed mass for two representative
clusters, which also happen to be the first two in Table 1. In the case of NGC 104, 
DARKexp and King models fit about equally well, while NGC 1805 is much better fit with 
DARKexp. First, we note that we use half-mass radii instead of core radii
because the latter are not defined
for DARKexp models. Further, because the outer-most projected density profile of DARKexp 
falls off as $r^{-3}$, the mass enclosed scales as $M(<r)\propto\,r^{-1}$. This relatively 
slow fall-off makes "total" mass a radius-dependent quantity, and requires one to decide 
where to truncate the DARKexp to calculate the "total" mass. We chose to use the last 
NG06 data point. With these choices, the ratio of the half-mass radii of best-fit models 
for NGC 104 is $R_{DARKexp}/R_{King}=0.95$, and the ratio of total masses is
$M_{DARKexp}/M_{King}=1.0$. For NGC 1851, the corresponding values are both 0.77.

\section{Fitting 3D light distribution of GCc with DARKexp and King
models}\label{threeD}

\begin{figure*}
  \begin{center}
    \leavevmode
      \epsfxsize=15cm\epsfbox{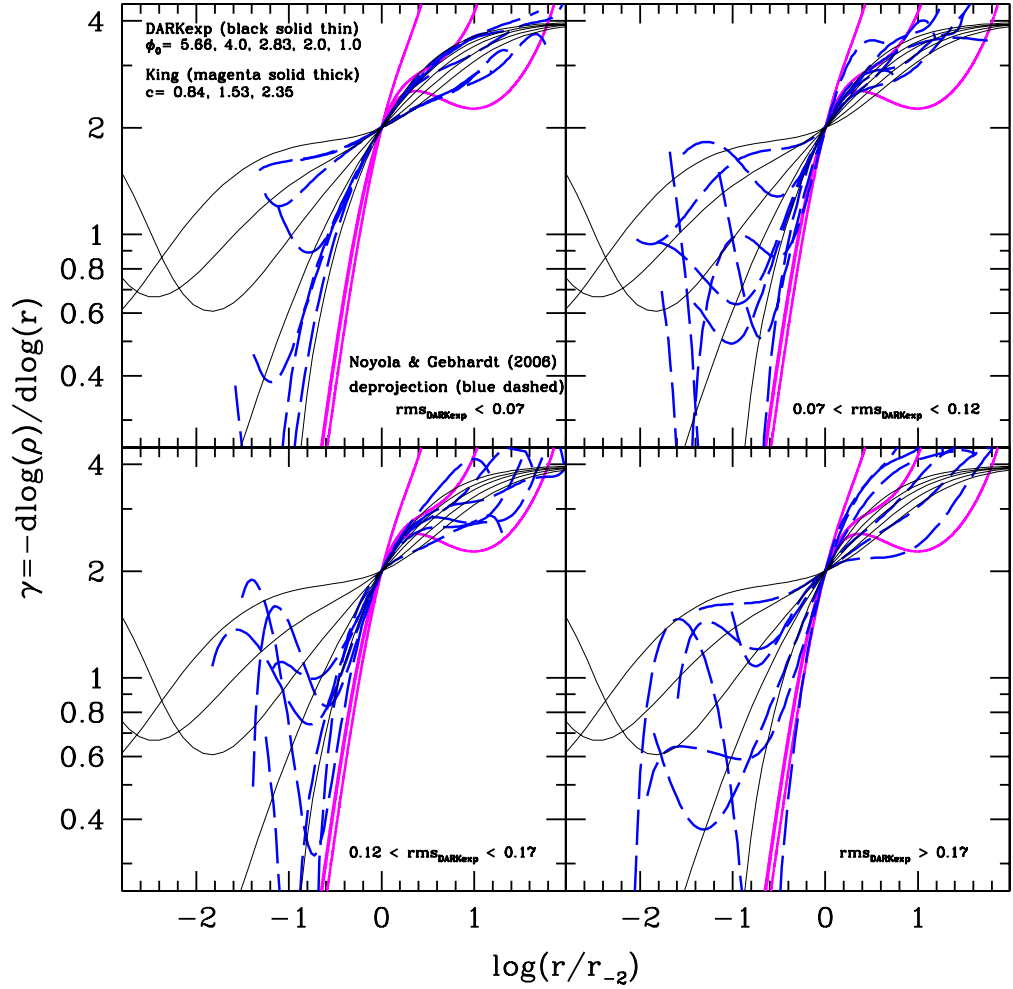}
\caption{3D log-log density profile slopes for 26 GCs deprojected by Noyola \& Gebhardt (2006) (blue dashed) 
compared to DARKexp (black) and King (magenta) models. All the curves are normalized to have
slope $\gamma=2$ at $r=r_{-2}$. The GCs are grouped into four panels based on how well they 
are fit by DARKexp; $rms$ ranges are shown in each panel.}
\label{slopesGCall}
  \end{center}
\end{figure*}

Of the 38 GC in the \cite{ng06} set, 12 show central dips in the
smooth non-parametric fits, and cannot be deprojected to yield 3D
density profiles. For the remaining 26 clusters, NG06 present their
non-parametric deprojections. We take these 26 GCs, smooth them to
eliminate the fluctuations due to the finite precision of their
published results (see the sixth paragraph of Section~\ref{twoD} above), 
and compute 3D density slope, $\gamma$. These are plotted in
Figure~\ref{slopesGCall}. For reference, we also plot five DARKexp and
four King models, with $\phi_0=5.66, 4.0, 2.83, 2.0, 1.0$ and
$c_K=0.84, 1.25, 1.83, 2.35$, respectively.  These particular values
for $\phi_0$ and $c_K$ were chosen arbitrarily.

Figure~\ref{slopesGCall} is another way of looking at the same GC
data. The 26 clusters are grouped by their DARKexp $rms$ values, with
the top left panel showing the 7 GCs which are fit by  DARKexp very
well, while the bottom right panel shows 6 GCs where DARKexp is a
relatively poorer fit.  Note that the bulk of the clusters have
shallow inner density slopes, though in most cases these are not as
shallow as those of King cores.  DARKexp, on the other hand, has a
range of slopes and slope derivatives at $r < r_{-2}$ that match GC well. 
At $r > r_{-2}$ some clusters are consistent with
DARKexp profiles, but a few show non-monotonic $\gamma$ behavior
reminiscent in character of King profiles.  Overall, even though
DARKexp provide better fits, similarities with King density profiles
are also seen at some radii.
 
\section{Conclusions}

\cite{k66} models, which are based on the assumption of two-body
collisional dynamics \citep{m96}, do not provide good fits to surface density
profile of many Globular Clusters, especially those whose data span a
large radial range. Deviations occur at small and large radii, which
are usually explained by physics beyond the King model.  Motivated by
these widespread departures, we ask if collisionless effects due to
collective behavior of cluster stars could be important.
In Section~\ref{coll}, we present two independent order-of-magnitude 
arguments that answer the question in the affirmative. 

We then use previously created Globular Cluster surface brightness
profiles to further test the relevance of collisionless effects.  Our
goal is made easier because there exist first-principles analytical
models of collisional and collisionless systems, the King-Madsen and
the DARKexp models, respectively.  Both were derived based on maximum
entropy statistical arguments. Both are one parameter families, and
can be readily converted to density profiles and hence compared to the
surface brightness profile of GCs (assuming GCs contain no dark matter
and have constant mass-to-light ratios).  We take the SB profiles from
the work of \cite{ng06} who present smooth non-parametric composite
profiles based on HST and ground-based data and fit these with King,
Madsen, and DARKexp density profiles. 

Our main finding is that DARKexp models fit considerably better than
either King or Madsen models, over the entire available radial range.
While this is not proof that collisionless dynamics has shaped the
mass distribution of GCs, it does suggest it as an interesting possibility.
It is already well known that the consequences of collisional processes, 
like the presence and hardening of binaries, mass segregation and evaporation 
of stars are important in observed and simulated clusters \citep{mh97}. 
In this paper we demonstrate that the overall density profile is well 
described by a collisionless prediction, the DARKexp family of models. 
Taken together, these observations suggest
that collisional and collisionless processes co-exist in GCs, but are
responsible for different sets of properties. An alternative
explanation is that a good DARKexp fit is purely fortuitous in that it
allows for a cuspy central profile and so phenomenologically accounts for
the effects of $M/L$ variation with radius as well as the effects of core collapse
or post core collapse, etc., but the physics is not related to collisionless relaxation;
DARKexp just happens to be a better fitting function to a very complex system.

Because King models are apparently not the best `typical' model for GC, we caution 
against using these to derive cluster global structural parameters, like characteristic 
radii and enclosed mass, especially if these are used in other types of analyses, for 
example, to look for correlations with metallicity, colors, age, etc. 

A secondary question we have addressed concerns an intriguing property of
self-gravitating systems. The derivation of Madsen and DARKexp models
is based on statistical mechanics; basically counting the number of 
particles in various states. In classical applications, all states have
large occupation numbers, so the Stirling approximation is commonly used. 
Madsen and DARKexp models depend critically on the accurate treatment of 
the low occupation number regime and so raise the question of how the DF 
should behave when  $n=1,2,3...$.  Should it be step-like, i.e. discrete, 
or should it be smooth?  \cite{m96} and \cite{k66} models
are the two corresponding versions. (DARKexp is smooth.) A comparison
of Madsen and King models for a subset of GCs shows the latter fit better. 

In the future it will be interesting to extend our analysis to include the
dynamical information on Globular Clusters. Velocity dispersion profiles,
though of poorer quality and of shorter radial range than surface brightness
profiles, are available for some clusters, and can be used in combination 
with the surface brightness profiles to assess the state of the GCs \citep{z12}.

Since DARKexp has already been shown to fit the results of collisionless
dark matter $N$-body simulations quite well \citep[both in the density,
$\rho(r)$ and energy space, $N(E)$;][]{whw10}, it is interesting to
compare the best-fitting shape parameter, $\phi_0$, to those obtained
in this work for GCs. $N$-body generated dark matter halos have a narrow
range of $\phi_0$ around 4--5. GCs (excluding those that are poorly fit
by DARKexp) mostly have $\phi_0$ between 1 and 3, because at approriate
radii these DARKexp profiles have shallow, or flat density slopes. A few 
GCs have $\phi_0$ values between 3 and 8. It is unclear why these values 
of $\phi_0$ are preferred in either one of the systems.

$$            $$
The Dark Cosmology Centre is supported by the Danish National Research 
Foundation. LLRW thanks Evan Skillman for valuable suggestions.

\clearpage


\clearpage

\begin{table}
\caption{DARKexp \& King model best-fit parameters for 38 GCs}
\begin{tabular}{ccc|ccc|cccc}
\hline
{NGC} &  & & $\phi_0$ & $c_{D13}$ & $rms_{D}$ (mag/sq.arcsec) & $\frac{\Psi(0)}{\sigma^2}$ & $c_{K}$ & $c_{K13}$ & $rms_{K}$ (mag/sq.arcsec) \\
\hline
  104 & 47 Tuc&  & 1.75 & 1.16 & 0.112 (0.050, 0.101) &  8.6 & 2.01 & 1.47 & 0.102 (0.011, 0.102) \\
 1851 &       &  & 2.25 & 1.35 & 0.062 (0.030, 0.054) &  9.6 & 2.26 & 1.78 & 0.693 (0.331, 0.609) \\
 1904 & M 79  &c?& 1.50 & 1.08 & 0.018 (0.005, 0.018) &  7.8 & 1.77 & 1.16 & 0.088 (0.059, 0.065) \\
 2298 &       &  & 0.30 & 0.80 & 0.044 (0.041, 0.015) &  7.1 & 1.56 & 0.86 & 0.047 (0.042, 0.022) \\
 2808 &       &  & 0.05 & 0.75 & 0.441 (0.084, 0.433) &  8.3 & 1.92 & 1.36 & 0.594 (0.110, 0.584) \\
 5272 & M 3   &  & 0.05 & 0.75 & 0.211 (0.052, 0.205) &  8.8 & 2.06 & 1.54 & 0.631 (0.161, 0.610) \\
 5286 &       &  & 1.00 & 0.94 & 0.156 (0.111, 0.109) &  7.6 & 1.71 & 1.07 & 0.175 (0.134, 0.112) \\
 5694 &       &  & 1.75 & 1.16 & 0.076 (0.043, 0.062) &  8.4 & 1.95 & 1.40 & 0.163 (0.053, 0.154) \\
 5824 &       &  & 3.00 & 1.75 & 0.134 (0.063, 0.118) & 10.1 & 2.37 & 1.89 & 0.782 (0.414, 0.663) \\
 5897 &       &  & 0.05 & 0.75 & 0.224 (0.066, 0.214) &  4.2 & 0.87 & 0.45 & 0.076 (0.039, 0.065) \\
 5904 & M 5   &  & 0.05 & 0.75 & 0.356 (0.041, 0.354) &  8.0 & 1.83 & 1.24 & 0.300 (0.045, 0.297) \\
 6093 & M 80  &  & 0.75 & 0.89 & 0.162 (0.056, 0.152) &  8.5 & 1.98 & 1.44 & 0.344 (0.096, 0.331) \\
 6205 & M 13  &  & 0.05 & 0.75 & 0.517 (0.105, 0.506) &  7.3 & 1.62 & 0.94 & 0.366 (0.066, 0.360) \\
 6254 & M 10  &  & 0.05 & 0.75 & 0.199 (0.040, 0.195) &  7.1 & 1.56 & 0.86 & 0.146 (0.028, 0.143) \\
 6266 & M 62  &c?& 1.75 & 1.16 & 0.073 (0.036, 0.064) &  7.8 & 1.77 & 1.16 & 0.122 (0.096, 0.075) \\
 6284 &       &c & 4.75 & 2.75 & 0.059 (0.042, 0.042) &  9.8 & 2.30 & 1.82 & 0.171 (0.143, 0.095) \\
 6287 &       &  & 0.50 & 0.83 & 0.106 (0.080, 0.070) &  6.7 & 1.44 & 0.74 & 0.118 (0.092, 0.074) \\
 6293 &       &c & 5.50 & 3.06 & 0.050 (0.030, 0.040) & 10.5 & 2.45 & 1.98 & 0.205 (0.169, 0.116) \\
 6333 & M 9   &  & 0.05 & 0.75 & 0.102 (0.027, 0.098) &  7.3 & 1.62 & 0.94 & 0.180 (0.010, 0.180) \\
 6341 & M 92  &  & 0.05 & 0.75 & 0.348 (0.051, 0.344) &  8.1 & 1.86 & 1.28 & 0.299 (0.039, 0.296) \\
 6352 &       &  & 2.50 & 1.46 & 0.101 (0.049, 0.089) & 11.2 & 2.59 & 2.11 & 0.134 (0.031, 0.131) \\
 6388 &       &  & 1.25 & 1.01 & 0.062 (0.030, 0.054) &  9.0 & 2.12 & 1.61 & 0.455 (0.154, 0.428) \\
 6397 &       &c & 3.50 & 2.09 & 0.120 (0.065, 0.100) &  9.7 & 2.28 & 1.80 & 0.312 (0.224, 0.217) \\
 6441 &       &  & 1.25 & 1.01 & 0.014 (0.006, 0.012) &  8.1 & 1.86 & 1.28 & 0.126 (0.036, 0.121) \\
 6528 &       &  & 3.00 & 1.75 & 0.167 (0.088, 0.142) &  9.1 & 2.14 & 1.64 & 0.057 (0.035, 0.045) \\
 6535 &       &  & 3.75 & 2.25 & 0.095 (0.075, 0.059) &  9.7 & 2.28 & 1.80 & 0.266 (0.236, 0.123) \\
 6541 &       &c?& 2.25 & 1.35 & 0.240 (0.094, 0.221) &  9.2 & 2.17 & 1.67 & 0.432 (0.192, 0.387) \\
 6624 &       &c & 3.25 & 1.92 & 0.151 (0.088, 0.123) &  9.1 & 2.14 & 1.64 & 0.151 (0.106, 0.107) \\
 6626 & M 28  &  & 1.50 & 1.08 & 0.049 (0.034, 0.035) &  7.9 & 1.80 & 1.20 & 0.077 (0.042, 0.064) \\
 6637 & M 69  &  & 0.05 & 0.75 & 0.052 (0.028, 0.044) &  7.0 & 1.53 & 0.82 & 0.057 (0.029, 0.049) \\
 6652 &       &  & 4.75 & 2.75 & 0.144 (0.076, 0.122) &  9.6 & 2.26 & 1.78 & 0.188 (0.089, 0.166) \\
 6681 & M 70  &c & 7.50 & 3.90 & 0.065 (0.042, 0.049) &  9.9 & 2.33 & 1.85 & 0.194 (0.121, 0.152) \\
 6712 &       &  & 0.05 & 0.75 & 0.099 (0.019, 0.098) &  5.8 & 1.20 & 0.59 & 0.037 (0.017, 0.033) \\
 6715 & M 54  &  & 2.00 & 1.25 & 0.104 (0.046, 0.094) &  9.4 & 2.21 & 1.72 & 0.596 (0.218, 0.554) \\
 6752 &       &c & 2.50 & 1.46 & 0.105 (0.039, 0.098) &  9.5 & 2.24 & 1.75 & 0.318 (0.087, 0.306) \\
 7078 & M 15  &c & 4.00 & 2.39 & 0.179 (0.048, 0.173) & 10.9 & 2.53 & 2.06 & 0.669 (0.320, 0.587) \\
 7089 & M 2   &  & 0.05 & 0.75 & 0.143 (0.036, 0.139) &  8.0 & 1.83 & 1.24 & 0.252 (0.031, 0.250) \\
 7099 & M 30  &c & 3.75 & 2.25 & 0.290 (0.078, 0.280) & 10.3 & 2.41 & 1.94 & 0.414 (0.170, 0.378) \\
\label{table1}
\end{tabular}
\end{table}

\label{lastpage}

\end{document}